\documentclass[letterpaper]{article}

\usepackage[T1]{fontenc}
\usepackage{microtype}
\usepackage{geometry}
\usepackage{setspace}
\usepackage{booktabs}

\usepackage[style = chem-acs]{biblatex}
\usepackage{graphicx}
\usepackage{placeins}
\usepackage{float}
\newfloat{scheme}{htbp}{los}
\floatname{scheme}{Scheme}
\floatname{chart}{Chart}
\newfloat{graph}{htbp}{loh}
\usepackage{amssymb}
\usepackage{amsmath}

\usepackage{chemformula}

\usepackage[version = 4]{mhchem}

\usepackage{authblk}
\author[1]{Ashish Kumar Singh}
\author[2, 3]{Irina V. Lebedeva}
\affil[1]{Institute of Atomic Physics and Spectroscopy, University of Latvia, Riga, Latvia}
\affil[2]{CIC nanoGUNE BRTA, Avenida de Tolosa 76, San Sebasti\'an 20018, Spain}
\affil[3]{Simune Atomistics, Avenida de Tolosa 76, San Sebasti\'an 20018, Spain}

\title{Trust-Limited Machine-Learning-Accelerated Nudged Elastic Band Calculations for Small-Polaron Hopping in Rutile TiO$_2$}

\date{*Email: kumar24@lu.lv, liv\_ira@hotmail.com}

\begin{document}

\twocolumn[{%
\maketitle

\begin{abstract}
Small-polaron hopping couples nuclear rearrangement to a localized, self-consistent electronic state, making first-principles nudged elastic band (NEB) optimization computationally demanding. We develop a trust-limited Gaussian-process-assisted NEB workflow and apply it to electron small-polaron migration in rutile TiO$_2$ using spin-polarized SIESTA DFT+$U$. A joint energy--force Mat\'ern Gaussian process guides band relaxation, while root-mean-square and maximum-atom displacement limits trigger additional DFT evaluations. Final pathways are accepted only after full-band verification of the projected DFT NEB forces. Three accepted 96-atom $[001]$ calculations with 11, 13, and 15 images yield migration barriers of 0.1645, 0.1637, and 0.1640~eV, respectively, with a spread of 0.793~meV. A separately calibrated local GP is activated during the 15-image calculation to refine the central transfer region. In the matched 96-atom, 13-image comparison, GP assistance reduces geometry-level DFT evaluations from 255 to 47 (81.6\%), while reproducing the conventional NEB barrier. For a 324-atom cell, the accepted $[001]$ pathway yields 0.1302~eV, whereas two closely matched $[111]$ runs give 0.1826 and 0.1828~eV. All accepted pathways meet a projected DFT NEB force criterion of 0.050~eV/\AA{}. A separate 324-atom, 13-image $[001]$ attempt remains unaccepted and is reported as such. These results demonstrate GP-guided polaron path optimization with explicit first-principles convergence assessment.
\end{abstract}
\vspace{0.7cm}
}]

\section{Introduction}

Small polarons arise from the localization of charge carriers through their coupling to lattice distortions, resulting in quasiparticles whose migration involves both charge transfer and local structural reorganization \cite{Holstein1959,Holstein1959a,Emin1975,Austin1969,Franchini2021}. Small-polaron formation and transport occur in a broad range of materials, including transition-metal oxides, ionic crystals, and molecular or organic systems \cite{Shluger1993,Franchini2021,Ruhle2011,Giannini2018}. In transition-metal oxides, localized electronic states have been investigated in materials such as CeO$_2$, $\alpha$-MoO$_3$, hematite, and SrTiO$_3$ \cite{Sun2017,Ding2014,Adelstein2014,Janotti2014}; localized hole carriers are also relevant to wide-gap oxides and ferroelectric interfaces \cite{Lany2009,Korbel2025}. The interaction between localized charge carriers and their surrounding lattice can strongly influence electronic transport, defect energetics, and charge-transfer processes. Consequently, determining the energetics and migration pathways of polarons is relevant to understanding the electronic properties of diverse functional materials.

Polaron formation and transport can significantly influence the performance of energy materials and electronic devices. Solar-energy conversion and photocatalysis are important applications of metal-oxide materials \cite{Fujishima1972,Regan1991}, where charge trapping and small-polaron formation can affect carrier recombination kinetics and interfacial charge transfer \cite{Rettie2016,Yang2013,Henderson2011,Craddock2026}. In intercalation electrodes, localized-carrier hopping can limit intrinsic electronic conductivity \cite{Maxisch2006,Luong2022}; small-polaron transport has also been investigated in different phosphate-electrode structures and compositions \cite{Banday2017,Sharma2020}. Charge localization and defect-mediated electronic transport are also relevant to resistive-switching devices \cite{Waser2007,Waser2009,Sawa2008,Talin2025}, while coupled ionic and electronic transport plays an important role in solid-oxide electrochemical systems \cite{Adler2004,Goodenough2000}. Despite their different functional roles, these materials share a computational challenge: determining the energetics and migration pathways of localized charge carriers. Methods that reduce the number of expensive first-principles evaluations while preserving migration-path accuracy are therefore of broad interest.

First-principles calculations are widely employed to investigate polaron localization, structural relaxation, and hopping barriers in oxides and related materials \cite{Setvin2014,Deskins2007,Bylaska2018,Franchini2021,Delile2019,Himmetoglu2014}. Comparisons across TiO$_2$ polymorphs have demonstrated that the stability and spatial extent of localized states depend on crystal structure and the treatment of self-interaction \cite{Elmaslmane2018}. Surface calculations likewise show that local coordination and the surrounding electrostatic environment affect polaron trapping and motion \cite{Reticcioli2018,Reticcioli2022}. Both electron and hole hopping pathways have been treated using first-principles approaches in TiO$_2$ \cite{Deskins2007,Carey2021}. 
Candidate migration pathways are commonly initialized by interpolation between initial and final configurations, followed by nudged elastic band (NEB) optimization to determine minimum-energy paths and migration barriers \cite{Jonsson1998,Henkelman2000a,Maxisch2006,Adelstein2014,Morita2023}. Site-localized electronic constraints can also be employed to investigate charge-transfer energetics and distinguish competing localized states \cite{Morita2023,Wu2018}. Molecular dynamics simulations probe finite-temperature carrier motion more directly \cite{Ahart2026,Wang2016,Birschitzky2025}, but generally require substantially greater statistical sampling to characterize infrequent hopping events.

The NEB method is a standard approach for determining minimum-energy paths (MEPs) and saddle-point barriers between known end states \cite{Henkelman2000,Henkelman2000a,Jonsson1998}. Its practical convergence can depend on the optimization algorithm and the initialization of the path, motivating improved optimization and path-correction protocols \cite{Herbol2017,Ruttinger2022}. In a conventional first-principles NEB calculation, energies and forces are repeatedly evaluated for every moving image while the band is optimized. For large supercells, this coupled iteration becomes expensive even when each individual electronic-structure calculation is tractable. Gaussian-process regression (GPR) offers a natural acceleration strategy because a probabilistic surrogate can be trained directly on energies and derivatives, used to relax an approximate path, and refined selectively where its uncertainty is large \cite{Rasmussen2006,Koistinen2017,Torres2019}.

Koistinen and co-workers developed GP-accelerated NEB schemes using reference energies and forces, uncertainty-directed image selection, and early stopping of surrogate relaxation \cite{Koistinen2017}. Their later formulation introduced inverse interatomic-distance measures and additional geometric stopping criteria, with final convergence assessed using accurate reference forces \cite{Koistinen2019}. Garrido Torres et al. developed uncertainty-driven ML-NEB and demonstrated it on both model systems and realistic DFT reactions, including molecular rearrangement and water dissociation \cite{Torres2019}. Other GP-based MEP and transition-state optimizers likewise reduce expensive gradient evaluations \cite{Denzel2019}. Consequently, earlier work is not restricted to the two-dimensional M\"{u}ller--Brown surface, and geometric trust control or reference-force verification cannot be claimed individually as new here.

More recent developments further broaden the landscape of surrogate-assisted path searches. Teng et al. combined physically motivated GP prior means with climbing-image NEB for gas-phase, interfacial, and bulk reactions \cite{Teng2024}. Onyango et al. implemented an on-the-fly energy--force GP calculator that adapts reference sampling to predictive uncertainty during NEB calculations \cite{Onyango2025}, while Simon et al. developed an active-learning NEB algorithm for identifying transition states in molecular and high-dimensional benchmark systems \cite{Simon2026}. A related but distinct approach trains machine-learning force fields through active learning over reaction pathways and subsequently checks the resulting barriers against DFT \cite{Schaaf2023}. These contributions, together with recent methodological reviews \cite{Goswami2026}, demonstrate that uncertainty-directed acquisition, adaptive surrogate refinement, and reference-level validation are established ideas. Their application to polaron migration nevertheless requires attention to the self-consistent electronic solutions underlying each energy and force observation.

Recent work also demonstrates machine-learning-assisted automation directly in polaronic materials. Yalcin et al. introduced PolFlow, combining a force field with explicit polaron-site labels, automated DFT localization checks and recovery, and active-learning exploration of defect--polaron configurations on TiO$_2$(110) \cite{Yalcin2026}. Its ConfML component uses ML-predicted configuration energies and simulated annealing to identify favorable arrangements of defects and localized polarons. These configurational moves do not determine the intervening minimum-energy migration path or its hopping barrier. The present work addresses this complementary task: GP-assisted optimization of an elastic band connecting localized polaron endpoints, with particular attention to central-region sampling, prediction errors, and full-band DFT force convergence.

Small-polaron hopping adds electronic-state sensitivity to this established optimization problem. On an analytic benchmark surface, each geometry has a well-defined energy and gradient. In first-principles calculations of polarons, electronic energies and forces are obtained through an iterative self-consistent-field (SCF) procedure, whose converged solution can depend on the initial electronic configuration and small changes in atomic positions. Metastable site and orbital occupations are documented challenges in localized-state calculations \cite{AllenWatson2014,Reticcioli2022}. This sensitivity can arise with semilocal DFT, DFT+$U$, and hybrid functionals, particularly when competing localized electronic states are present; the balance between localization and delocalization is itself sensitive to the exchange--correlation treatment \cite{Lany2009,Elmaslmane2018}. Near the charge-transfer region, donor- and acceptor-localized states may converge to distinct electronic solutions at similar nuclear geometries, potentially introducing discontinuities or inconsistencies in the energy and force data used to train a surrogate model. Localization changes during DFT calculations are also a central concern in automated polaron workflows \cite{Yalcin2026}. A smooth GP fit does not, by itself, establish that a consistent electronic branch has been sampled. This motivates restricting surrogate excursions and carefully resolving the central region of the migration pathway. Demonstrating electronic-branch switching in a particular trajectory, however, requires converged site-resolved charge populations or spin densities and, ideally, repeated electronic initializations at the same geometry. SCF input settings alone are insufficient to establish such switching.

We therefore formulate GP-assisted NEB as a sequence of trust-limited surrogate updates. Root-mean-square (RMS) and maximum-atom displacement criteria bound surrogate motion relative to evaluated structures, and additional DFT evaluations refine the band when these limits are reached. Final acceptance is based on full-band DFT forces. A separately calibrated local GP is also tested for the central transfer region in the 96-atom, 15-image calculation; all other accepted production calculations use the global GP alone.

In this work, we demonstrate the proposed trust-limited GP-assisted NEB framework using electron small-polaron hopping in rutile TiO$_2$ as a representative case study. Spin-polarized DFT+$U$ calculations with SIESTA are performed along $[001]$ for different NEB discretizations and supercell sizes, and an additional 324-atom $[111]$ pathway probes a distinct hopping geometry. A direct comparison with conventional NEB for the 96-atom, 13-image $[001]$ case is used to quantify the reduction in first-principles evaluations; no conventional 324-atom NEB baseline is reported. All accepted GP-assisted barriers are independently verified against first-principles NEB forces. Although demonstrated here for rutile TiO$_2$, the framework is not intrinsically restricted to this material and may be extended to other systems exhibiting localized-carrier hopping, provided that consistent electronic states can be obtained along the migration pathway.

\section{Methods}

\subsection{Electronic-structure calculations and rutile TiO$_2$ models}

Spin-polarized DFT+$U$ calculations were performed using the SIESTA code~\cite{Soler2002, Garcia2020}. Exchange-correlation interactions were treated within the generalized gradient approximation (GGA) using the Perdew--Burke--Ernzerhof (PBE) functional~\cite{Perdew1996}. Optimized Norm-Conserving Vanderbilt (ONCVPSP) fully relativistic pseudopotentials~\cite{Hamann2013} in the PSML format~\cite{Garcia2018} were sourced from the PseudoDojo database~\cite{vanSetten2018}. The pseudopotential for Ti explicitly included the $3s$ and $3p$ semicore states in the valence pool, corresponding to a valence configuration of $3s^2 3p^6 3d^2 4s^2$. Valence states were described using a double-$\zeta$ polarized (DZP) basis set, with spatial confinement defined by an energy shift of 0.085~eV. This energy shift was selected based on convergence tests performed at the midpoint geometry of the linearly interpolated hopping path, using equal site-resolved spin populations on the two Ti cations defining the hop as the target criterion. A real-space grid cutoff of 500~Ry was employed. For all NEB-image calculations, electronic self-consistency was evaluated via convergence of the Hamiltonian and density matrix, with tolerances set to 10$^{-3}$~eV ($dH_{\max}$) and 10$^{-4}$, respectively, accelerated using Pulay mixing with a 5-step history and a weight of 0.1. 

To mitigate self-interaction-induced delocalization of Ti $3d$ electrons, an effective Hubbard correction of $U_{\mathrm{eff}}=U-J=4.0$~eV was applied to the Ti $3d$ orbitals using the Dudarev formulation \cite{Dudarev1998}. The Hubbard parameter was selected through a systematic localization-based convergence study performed on the excess-electron polaron structures. Different values of $U_{\mathrm{eff}}$ were tested for the relaxed initial and final polaron configurations, as well as for the midpoint geometry obtained by linear interpolation between the endpoints. The spatial distribution of the excess electron was assessed using site-resolved Hirshfeld and Voronoi population analyses. The lowest tested value yielding well-defined charge localization in all three configurations was $U_{\mathrm{eff}}=4.0$~eV. This is a localization-based choice, rather than a linear-response determination of the Hubbard interaction \cite{Cococcioni2005}, and does not by itself establish convergence of the hopping barrier with respect to $U_{\mathrm{eff}}$. The selected value was retained throughout the subsequent NEB calculations to maintain a consistent description of the localized electronic states.

The selected Hubbard parameter is in close agreement with the value of 3.9~eV obtained for rutile TiO$_2$ using the constrained random-phase approximation (cRPA) \cite{Setvin2014}. Previous studies have also demonstrated the sensitivity of excess-electron localization and its accompanying lattice relaxation in TiO$_2$ to the Hubbard correction \cite{Morgan2007,Morgan2010,Reticcioli2022}. Semilocal exchange--correlation functionals can favor excessive delocalization of Ti $3d$ electrons due to self-interaction errors, whereas the on-site Hubbard correction penalizes fractional orbital occupations and can stabilize localized polaronic states accompanied by lattice distortion. Hybrid exchange--correlation functionals provide an alternative treatment that can also support charge localization, but generally at a higher computational cost than DFT+$U$ \cite{Ding2014,Wu2018,Delile2019,Franchini2021}.

Two supercell sizes were constructed to model electron small-polaron hopping along {$[001]$}: a $2\times2\times4$ rutile supercell (96~atoms) and a $3\times3\times6$ supercell (324~atoms). The 96-atom system served as the primary benchmark for algorithmic validation and path-discretization sensitivity, whereas the 324-atom system was used to evaluate the scalability of the GP-assisted workflow and assess cell-size effects on the hopping barrier. An additional $[111]$ pathway was evaluated in the same 324-atom cell using 13 images. For the 324-atom input geometry, the $[001]$ endpoints are Ti157 and Ti163, and the $[111]$ endpoints are Ti157 and Ti302 (1-based input-file indices). The local spin-density illustrations retain their source-figure labels Ti53 and Ti55; these labels should not be interpreted as 324-atom input-file indices. 
A $2\times2\times2$ Monkhorst--Pack $k$-point grid \cite{Monkhorst1976} was employed for the 96-atom cell, while $\Gamma$-only sampling was used for the 324-atom supercell. Because the sampling schemes differ, variations between the two cells reflect both finite-size and $k$-point integration effects rather than spatial confinement alone.

The structures were prepared, optimized, and analyzed using the Atomistic Simulation Advanced Platform (ASAP) package~\cite{Marchesin2022, SIMUNE2025}. For each localized-polaron endpoint, one excess electron was added to the supercell (net charge $-1$) and initialized on the target Ti site. No constraints were applied to the collinear spin. The initial and final endpoint structures were relaxed independently. Ionic relaxations were performed until maximum residual atomic forces were below 0.005~eV/\AA{}. In the prepared $[111]$ endpoint geometry, the Ti157--Ti302 displacement is $(2.300,2.300,1.480)$~\AA{}, corresponding to $(a/2,a/2,c/2)$ and a separation of 3.573~\AA{}. Both localized endpoint structures were relaxed before construction of the interpolated band.

\begin{figure*}[!t]
    \centering
    \includegraphics[width=0.75\textwidth]{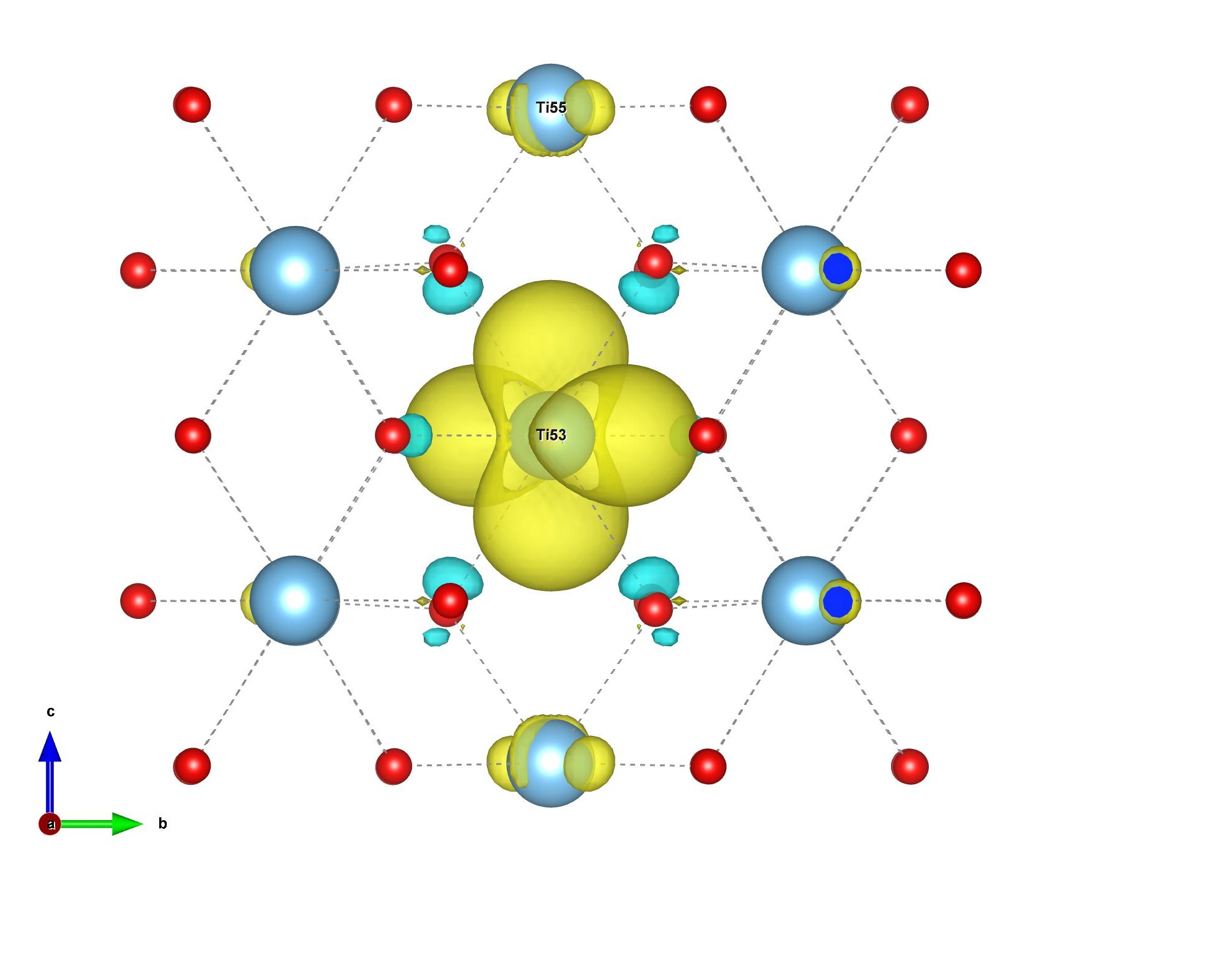}
    \caption{Local atomic structure and spin-density
    distribution illustrating an electron small polaron localized
    at Ti53 in rutile TiO$_2$. The rendering displays only
    the local environment around the $[001]$ pair labelled
    Ti53--Ti55 in the source visualization.
    Yellow and cyan isosurfaces indicate positive and
    negative spin-channel differences
    [Eq.~\eqref{eq:spin_density}], respectively.
    Ti and O atoms are represented by light-blue
    and red spheres, respectively.}
\label{fig:polaron_structure}
\end{figure*}

Figure~\ref{fig:polaron_structure} illustrates a local atomic environment and spin-density distribution representative of a $[001]$ small-polaron hop. Its Ti53/Ti55 labels follow the numbering in the visualization; they are not used to specify the endpoints of the 324-atom NEB production calculations.

The spatial spin density is evaluated from the spin-channel electron density distribution as
\begin{equation}
\Delta\rho_s(\mathbf{r}) =
\Delta\rho_\uparrow(\mathbf{r}) -
\Delta\rho_\downarrow(\mathbf{r}),
\label{eq:spin_density}
\end{equation}
\noindent
where $\Delta\rho_\uparrow(\mathbf{r})$ and $\Delta\rho_\downarrow(\mathbf{r})$ represent the differences between the self-consistent and atomic reference electron densities for the spin-up and spin-down channels, respectively. For an unpolarized atomic reference, $\Delta\rho_s(\mathbf{r})$ reduces directly to the total spin density, $\rho_\uparrow(\mathbf{r}) - \rho_\downarrow(\mathbf{r})$.

The yellow and cyan isosurfaces represent positive ($\Delta\rho_s>0$) and negative ($\Delta\rho_s<0$) spin polarization, respectively. The dominant positive spin polarization is concentrated around Ti53, indicating localization of the excess electron at this site. Smaller regions of opposite spin polarization are observed around neighbouring atoms. The illustrated Ti53--Ti55 pathway follows the crystallographic $c$ axis and represents the $[001]$ electron-hopping direction.

The 324-atom ($3\times3\times6$) GP-assisted NEB calculation was performed on the Hyperion supercomputer at the DIPC Supercomputing Center, using SIESTA version 5.2.0. The calculation was executed on a compute node equipped with Intel Xeon Gold 6342 (Ice Lake) processors, providing 48 CPU cores per node, of which 24 were allocated to the calculation.

\subsection{Nudged elastic band formulation}
For image $i$, the physical force is $\mathbf{F}_i=-\nabla E(\mathbf{R}_i)$, and $\boldsymbol{\tau}_i$ is the normalized path tangent, evaluated using the improved tangent prescription of Henkelman and J\'onsson \cite{Henkelman2000a}. For equal spring constants, the NEB force is
\begin{equation}
\label{nebf}
\mathbf{F}_i^{\mathrm{NEB}} = \mathbf{F}_i - (\mathbf{F}_i \cdot \boldsymbol{\tau}_i) \boldsymbol{\tau}_i + k_{\mathrm{s}}(d_{i,+}-d_{i,-})
\boldsymbol{\tau}_i,
\end{equation}
where $d_{i,+}$ and $d_{i,-}$ are adjacent-image separations in the path coordinate representation. The climbing-image mode was disabled for all runs. Equation~\eqref{nebf} including the tangential spring contribution therefore applies to all moving images.

Final acceptance requires
\begin{equation}
\label{neb_convergence}
\max_{i \in \mathcal{I}_{\mathrm{mov}}}
\max_a
\left\|
\mathbf{F}_{i,a}^{\mathrm{NEB,DFT}}
\right\|_2
<
0.050\,\mathrm{eV}/\text{\AA},
\end{equation}
where $\mathcal{I}_{\mathrm{mov}}$ denotes the moving images, $a$ indexes atoms, and $\mathbf{F}_{i,a}^{\mathrm{NEB,DFT}}$ is the three-component atomic contribution to the NEB force evaluated using DFT energies and forces.

The 96--11--G run used FIRE \cite{Bitzek2006}, while the 96--13--G, 96--15--GL, 324--11--G, and both 324--13--G attempts used MDMin as implemented in the Atomic Simulation Environment (ASE) \cite{Larsen2017}. The runs used a spring constant of $0.4\,\mathrm{eV}/\text{\AA}^{2}$, a time step of 0.01, and a maximum optimizer step of $0.004\,\text{\AA}$.

Energy profiles are plotted against cumulative image-to-image displacement normalized to the interval $[0,1]$, with all energies reported relative to the initial relaxed endpoint. Relative energy differences between the initial and final endpoints are fully preserved. Crucially, neither a stationary barrier profile nor a low surrogate force residual alone is sufficient to guarantee a converged DFT minimum energy path.

\subsection{Training configurations and Gaussian process prior}
Let $A$ contain $N_a$ atoms, $D=3N_a$ their Cartesian degrees of freedom, and $q \in \mathbb{R}^D$ their coordinates. The retained dataset contains $N_g$ evaluated geometries, each contributing the full-system DFT energy and their Cartesian forces:
\begin{equation} \label{D}
\mathcal{D} = \left\{ \left( q_n, E_n, F_{n, A} \right) \right\}_{n=1}^{N_g}, \quad M = N_g (1 + D)
\end{equation}

The joint energy--force GP formulation for the RMS Cartesian metric and Mat\'{e}rn covariance kernel is defined as follows~\cite{Koistinen2017, Koistinen2019, Rasmussen2006}. Their respective derivatives assume that the periodic-image mapping remains locally fixed along the path. The corresponding covariance derivatives, prediction formulas, and numerical treatment are provided in the Supplementary Information.
\begin{equation} \label{Eq}
E(q)\sim \mathcal{GP}\left(m(q), k(q, q')\right), \quad F(q) = -\nabla_q E(q)
\end{equation}

For a constant prior energy mean $m = E_0$, the corresponding prior force mean vanishes identically ($\nabla_\mathbf{R} E_0 = \mathbf{0}$). This prior mean $E_0$ serves solely to eliminate an arbitrary baseline energy shift without altering the predicted forces, and it remains distinct from the physical endpoint energy reference utilized for relative profile plotting. With $\Delta_j=q_j-q_j'$ denoting displacement of Cartesian component $j$ under a locally fixed periodic-image mapping, the covariance is
\begin{equation} \label{eq:rho_and_k}
\begin{aligned}
    \rho^2 &= \frac{1}{N_a}\sum_{j=1}^{D}\Delta_j^2,
    &u &= \frac{\sqrt{5}\rho}{\ell},\\[0.6em]
    k &= \sigma_f^2\left(1+u+\frac{u^2}{3}\right)e^{-u}.
\end{aligned}
\end{equation}
Here, $\rho$ is the atomic-RMS coordinate separation, $\ell$ is the correlation length, and $\sigma_f$ is an energy scale. The metric distance is normalized by the total number of atoms ($N_{\text{a}}$) rather than the total number of Cartesian degrees of freedom. This corresponds to a Mat{\'e}rn-$5/2$ covariance kernel~\cite{Rasmussen2006}. These Cartesian derivatives remain valid within any domain where the periodic-image mapping is locally constant. No differentiability is assumed across boundary crossings where the image mapping switches.

The Supplementary Information provides the joint energy--force covariance derivatives, posterior predictions and uncertainties, numerical linear-system formulation, training-data update expressions, and details of training-set retention and DFT evaluation-count accounting. The trust limits, acquisition decisions and full-band DFT acceptance protocol are described below.

\subsection{Cartesian coordinates and trust metrics}

All atoms are included in the surrogate coordinate representation and are allowed to move during surrogate optimization in the reported calculations. The GP is trained on the full-system DFT energies and Cartesian forces.

For the minimum-image displacement $\Delta\mathbf{R}_{a,n}=\mathbf{R}_a-\mathbf{R}_{a,n}$ of atom $a$ relative to retained DFT geometry $n$, define the per-reference displacement measures
\begin{equation} \label{eq:d_rms}
    d_{\mathrm{RMS}}(R,R_n) = \left[\frac{1}{N_a}\sum_{a=1}^{N_a}\|\Delta\mathbf{R}_{a,n}\|^2\right]^{1/2},
\end{equation}
\begin{equation} \label{eq:d_max}
    d_{\mathrm{max}}(R,R_n) = \max_{1\le a\le N_a}\|\Delta\mathbf{R}_{a,n}\|.
\end{equation}
The controller monitors proximity to previously evaluated configurations and displacement accumulated in the current optimization cycle using these RMS and maximum-atom measures. The RMS measure normalizes displacement by system size, while the maximum-atom measure guards against localized excursions. When the configured geometric limits are reached, surrogate relaxation stops and the corresponding configuration can be evaluated by DFT.

The controller alternates bounded surrogate motion and targeted DFT evaluation. A training-distance or cycle boundary can stop the optimizer before surrogate force convergence and trigger a reference acquisition. Acquisition logs distinguish these events from uncertainty or error-driven refinement and scheduled central-region sampling. The pre-acquisition posterior must be retained for an unbiased prediction audit before adding the new observation.

When a candidate band is submitted for first-principles verification, its image coordinates are held fixed and all moving-image DFT energies and forces are evaluated. The projected DFT NEB forces are then compared with the acceptance criterion in Eq.~\eqref{neb_convergence}.
Unsuccessful verification configurations are incorporated into the training data for subsequent surrogate refinement. The production workflow uses surrogate-force targets of 0.025~eV/\AA{} initially and 0.0125~eV/\AA{} during tightened refinement. The acceptance decision itself depends on the DFT-projected NEB forces, not on GP force predictions or uncertainty estimates. Details of training-set management and DFT evaluation accounting are given in the Supplementary Information.

\subsection{Calculation labels}
Calculation labels encode the total number of atoms, the total number of NEB images, and the prediction model architecture. G denotes a global GP model, whereas GL indicates a global GP augmented with an active central-region local predictor. The 324-atom, 13-image calculations are distinguished by explicit direction suffixes, 324--13--G$_{[001]}$ and 324--13--G$_{[111]}$, while labels without a direction suffix denote $[001]$ hopping. 

The training-distance RMS and maximum-atom thresholds for 96--11--G, 96--13--G, 96--15--GL, 324--11--G and 324--13--G$_{[111]}$ were set to 0.015/0.025, 0.015/0.015, 0.012/0.012, 0.015/0.015 and 0.015/0.015~\AA{}, respectively. The corresponding cycle-displacement RMS and maximum-atom limits were 0.030/0.020, 0.020/0.015, 0.015/0.015, 0.020/0.015 and 0.020/0.015~\AA{}, respectively. The unaccepted 324--13--G$_{[001]}$ trial used training-distance RMS/maximum limits of 0.015/0.015~\AA{} and per-cycle RMS/maximum limits of 0.020/0.015~\AA{}. The optimized global GP hyperparameters and retained-geometry capacity limits are summarized in Table~\ref{tab:gp_parameters}.

\subsection{Local GP treatment of the central polaron-transfer region}
The central charge-transfer region was sampled independently to support an optional local surrogate predictor. This localized sampling strategy was evaluated for central image~6 in run 96--13--G and central image~7 in run 96--15--GL. For 96--15--GL, auxiliary central sampling was triggered every three controller iterations whenever the RMS novelty metric exceeded 0.0015~\AA{}. The local predictor's RMS distance and maximum atomic displacement trust thresholds were set to 0.004~\AA{} and 0.010~\AA{}, respectively. The 324-atom calculations used the global GP alone. For 96--15--GL, local-GP activation followed eight new central-image DFT acquisitions, together with the original central seed geometry, giving nine locally stored configurations.

\begin{table*}[!t]
\centering
\caption{Calculation labels and global GP parameters. The dagger marks the additional 13-image $[001]$ run that was not accepted by the DFT NEB force criterion.}
\label{tab:gp_parameters}
\begin{tabular}{lcccccc}
\hline
Label & Atoms & Images & Active atoms & Global $\sigma_f$ (eV) & Global $\ell$ (\AA) & Retained cap \\
\hline
96--11--G  & 96  & 11 & 96  & 3.000 & 0.1200 & 40 \\
96--13--G  & 96  & 13 & 96  & 0.250 & 0.0400 & 45 \\
96--15--GL & 96  & 15 & 96  & 0.625 & 0.0550 & 40 \\
324--11--G & 324 & 11 & 324 & 0.300 & 0.0333 & 40 \\
324--13--G$_{[001]}^{\dagger}$ & 324 & 13 & 324 & 0.225 & 0.0333 & 40 \\
324--13--G$_{[111]}$ & 324 & 13 & 324 & 0.225 & 0.0300 & 40 \\
\hline
\end{tabular}
\end{table*}

The local model was calibrated via three-fold cross-validation performed on the eight acquired central-image geometries, employing a force-predictive length-scale optimization followed by uncertainty calibration. The posterior GP expressions apply identically to this localized training subset. The local GP interpolates DFT energies and forces at sampled configurations; like the global GP, it does not independently determine the electronic state. Its operation within the trust-limited controller is assessed here as part of the integrated workflow.

\subsection{Implementation}
The workflow is implemented in Python using ASE~\cite{Larsen2017} for atomic structure manipulation, NEB path management, and coupling to the electronic structure calculator. SIESTA~\cite{Soler2002, Garcia2020} evaluates the underlying spin-polarized DFT+$U$ energies and atomic forces within a localized atomic-orbital  framework. All GP regression routines, covariance matrix assembly, trust-distance evaluations, acquisition logic, training-set retention policies, DFT job orchestration, and prediction audits are executed directly by the custom workflow codebase.

\section{Results and Discussion}
\subsection{96-atom system: DFT benchmarking and central-region GP validation}\label{sec:conv_neb_comparison}

The 96-atom controller acquired repeated central-image configurations and recorded pre-acquisition force-prediction audits, motivating focused surrogate refinement near the polaron-transfer region. These diagnostics reveal localized difficulties in surrogate prediction; they do not, by themselves, identify distinct self-consistent electronic branches.

Furthermore, the global controller does not demand a fully converged surrogate pathway after each database update. Instead, surrogate path relaxation remains bounded by both the spatial distance to retained DFT training configurations and the cumulative displacement accumulated within the active surrogate cycle. Crossing either trust boundary halts surrogate minimization to trigger additional first-principles evaluations. This implementation adapts established early-stopping principles~\cite{Koistinen2017, Koistinen2019} to small-polaron migration pathways. Systematic comparison against looser trust thresholds remains necessary to independently quantify their individual impacts on search reliability and total computational cost.

This behavior is directly observed in the 96-atom production calculation. In one surrogate cycle, relaxation stopped after 27 optimizer steps at its prescribed atomic-displacement boundary even though the projected surrogate NEB force remained approximately 0.45~eV/\AA{}. A later cycle stopped at the training-distance boundary after 20 steps, triggering an additional DFT evaluation. These events illustrate the intended operating regime of the controller: reaching the geometric validity boundary is treated as an acquisition criterion, not as a failed optimization.

These events show that reaching the trust boundary, rather than complete surrogate force convergence, can determine when DFT sampling occurs.

A low surrogate NEB force does not, by itself, establish convergence of the underlying first-principles MEP. The production controller therefore separates surrogate convergence from DFT convergence. Once the GP band satisfies the surrogate force and uncertainty gates, the band coordinates are frozen and all internal images are reevaluated using DFT. The projected NEB forces are then reconstructed from DFT energies and forces at exactly those geometries.

The 96-atom calculation demonstrates why this distinction is necessary. An intermediate full-band verification produced a DFT barrier of 0.1688 eV, which was already close to the ultimately converged value. Nevertheless, the maximum projected DFT NEB force was 0.2210~eV/\AA{}, well above the required convergence threshold. The apparently reasonable barrier therefore corresponded to a band that was still substantially unconverged in force space. The verification was rejected and its first-principles data were incorporated into the GP before optimization continued.

Subsequent active learning and surrogate refinement ultimately produced a DFT-verified band with a barrier of 0.1645~eV and a maximum projected DFT NEB force of 0.0434~eV/\AA{}. Thus, the barrier changed by only a few meV between the failed and successful verification stages even though the maximum DFT NEB force decreased by approximately a factor of five. This provides a concrete demonstration that apparent convergence of the energy profile is not sufficient for establishing a first-principles MEP.

To assess the reduction in first-principles evaluations, a conventional DFT-based NEB calculation was performed for the 96-atom ($2\times2\times4$) rutile TiO$_2$ supercell using 13 images. The conventional and GP-assisted calculations employed identical DFT and SCF settings, NEB convergence criteria, and MDMin optimization parameters, with the climbing-image method disabled in both cases.

The conventional NEB calculation required 23 steps, including the initial full-band evaluation. With two fixed endpoints and 11 moving images, this corresponds to 13 initial DFT evaluations followed by
22 iterations involving 11 moving-image evaluations each, giving a total of 255 geometry-level DFT energy--force evaluations. The resulting migration barrier was 0.163778~eV.

In comparison, the corresponding GP-assisted calculation (96--13--G) required only 47 DFT evaluations, comprising 13 initial configurations, 23 subsequent DFT acquisitions, and 11 moving-image evaluations during final DFT verification. The resulting migration barrier was 0.1637~eV, with a maximum projected DFT NEB force of 0.0400~eV/\AA{}, satisfying the convergence threshold of 0.050~eV/\AA{}.

The two approaches yielded migration barriers differing by only approximately 0.1~meV. Under otherwise identical computational settings, the GP-assisted approach reduced the number of DFT evaluations from 255 to 47, corresponding to a reduction of 81.6\%, or approximately 5.43 times fewer first-principles evaluations. This comparison quantifies the savings in DFT evaluations rather than wall-clock time and does not include the computational overhead associated with GP training and prediction.

The three completed 96-atom {$[001]$} calculations use 11, 13 and 15 total images and give barriers of 0.1645, 0.1637, and 0.1640~eV. Their maximum projected DFT NEB forces are 0.0434, 0.0400 and 0.0346~eV/\AA{}, respectively, each below the common 0.050~eV/\AA{} criterion.

The barrier spread of 0.793 meV shows agreement across the tested setups. Because image count, GP parameters, trust limits and optimizer settings change together, this is a combined robustness test. A controlled image-number study would hold the other settings fixed. The observed spread is not an uncertainty estimate for the DFT approximation or for the infinite-cell barrier.

The accepted 96-atom bands were obtained with distinct controller settings rather than repeated runs with one parameter set. The 11-image calculation used FIRE, whereas the 13- and 15-image calculations used MDMin. The global GP parameters and retained training-set caps differ substantially (Table~\ref{tab:gp_parameters}), as do their trust limits (Methods). The 15-image calculation further used deliberately scheduled central-region acquisitions and an activated local predictor. It is therefore not a controlled test of image number or of local prediction in isolation.

The final full-band prediction audits provide an additional diagnostic beyond barrier reproducibility. For 96--13--G, the mean force RMSE was 0.0060~eV/\AA{} compared with a mean predicted RMS force uncertainty of 0.0342~eV/\AA{}. The corresponding values for 96--15--GL were 0.0045 and 0.0290~eV/\AA{}, respectively. These mean uncertainties exceed the observed mean errors at final verification, but do not establish calibrated uncertainty coverage at every image or throughout optimization. The results establish acceptance under distinct surrogate configurations, without attributing an independent evaluation-count improvement to a particular optimizer, trust limit, or local-GP choice.

The local model addresses the central transfer region, where the SCF solution may be particularly sensitive to geometry. In 96--13--G, the accepted band was obtained before the eight-acquisition local-model threshold was reached, so its production predictions remained global. In 96--15--GL, eight acquired image-7 configurations were available for three-fold force-prediction validation. The resulting length scale was $\ell=0.160$~\AA{} and the uncertainty-calibrated energy scale was $\sigma_f=1.0$~eV. Held-out central-image predictions gave a mean force RMSE of 0.0253~eV/\AA{}, a maximum atom-force error of 0.2837~eV/\AA{}, and force $z$-RMS of 1.267. The model was activated with nine stored central geometries, including the initial seed. The controller subsequently recorded local-GP predictions during production optimization. Direct localization diagnostics remain necessary to attribute individual errors to SCF branch changes.

After activation, the controller explicitly reported the central prediction model as the local GP and resumed production surrogate optimization. In the first local-GP-active cycle, the projected surrogate force decreased from 0.2922 eV/\AA{} to 0.0226 eV/\AA{} after 31 optimizer steps. Full-band DFT verification then showed that the candidate path was not yet converged: although the barrier was already 0.1646 eV, the true maximum projected DFT NEB force remained 0.0960 eV/\AA{}. The verification configurations were incorporated into the training data and the surrogate force limit was tightened.

With the local GP still active, the next surrogate refinement reduced the projected force to 0.0120 eV/\AA{}. A second full-band DFT verification then yielded a barrier of 0.1640 eV and a true maximum projected DFT NEB force of 0.0346 eV/\AA{}, satisfying the first-principles convergence criterion. The final full-band prediction audit gave mean force RMSE = 0.0045 eV/\AA{}, mean predicted RMS force uncertainty = 0.0290 eV/\AA{}, and an error/uncertainty ratio of 0.155. This calculation therefore provides the direct production demonstration that the crossing-region local GP can be calibrated, activated, and used within the global trust-limited controller while final acceptance remains determined by DFT NEB forces.

\begin{figure*}[!t]
    \centering
    \includegraphics[width=0.80\textwidth]
    {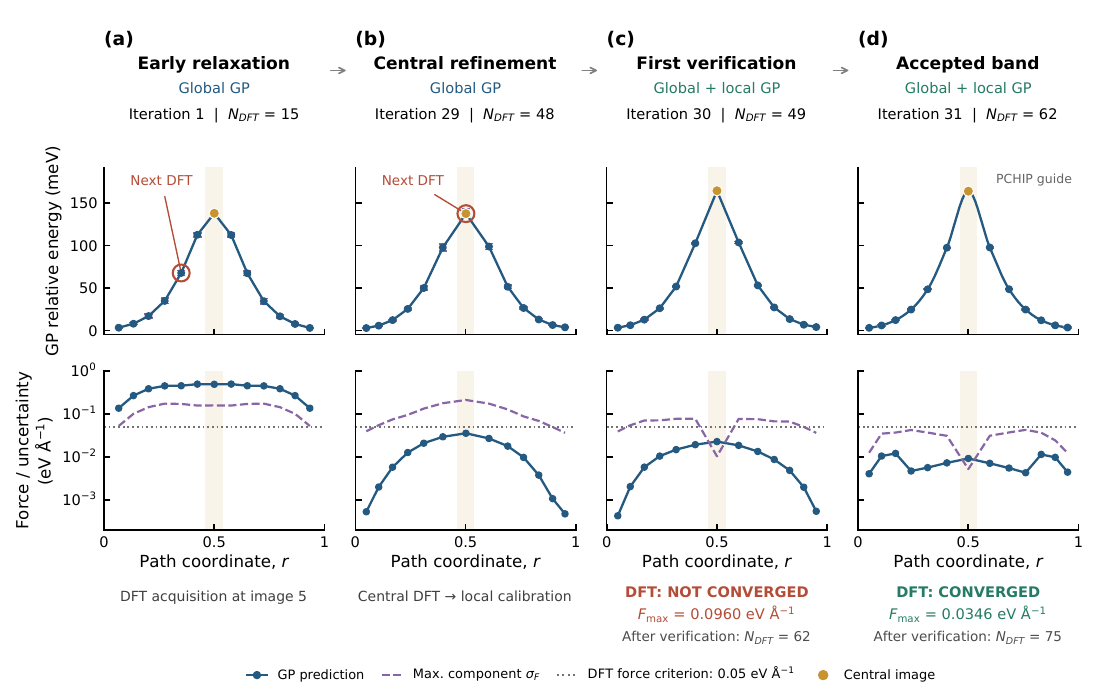}
    \caption{Evolution of the GP-assisted NEB optimization for the 96-atom rutile TiO$_2$ supercell using 15 images. The upper panels show the GP-predicted relative energy profiles, while the lower
    panels show the predicted NEB forces (blue) and maximum force-component uncertainties (purple dashed lines). The horizontal dotted line indicates the DFT NEB force acceptance criterion
    of 0.050~eV/\AA{}. (a) Early relaxation using the global GP, with the next DFT acquisition indicated. (b) Central-region refinement following repeated DFT sampling. (c) First full-band DFT verification following activation of the local GP, which failed the force criterion   ($F_{\max}=$~0.096~eV/\AA{}). (d) Final accepted band following additional refinement and DFT verification, with $F_{\max}=$~0.035~eV/\AA{}. The gold markers identify the central image,
    and the shaded regions highlight the central transfer region. The smooth interpolation in (d) is
    a visual guide.} \label{fig:polaron_evolution}
\end{figure*}

The successive stages of the 96--15--GL optimization are illustrated in Fig.~\ref{fig:polaron_evolution}. The initial global-GP relaxation and subsequent central-region refinement are shown in Figs.~\ref{fig:polaron_evolution}(a) and (b), respectively. Following activation of the calibrated local GP, the first full-band DFT verification failed to satisfy the force convergence criterion despite an apparently stabilized energy profile [Fig.~\ref{fig:polaron_evolution}(c)]. Further surrogate refinement and DFT verification produced an accepted pathway
[Fig.~\ref{fig:polaron_evolution}(d)]. These results demonstrate that the evolution of the surrogate energy profile and reduction of GP-predicted forces are insufficient to establish first-principles convergence without explicit DFT verification.

\begin{figure*}[!t]
    \centering
    \includegraphics[width=0.85\textwidth]
    {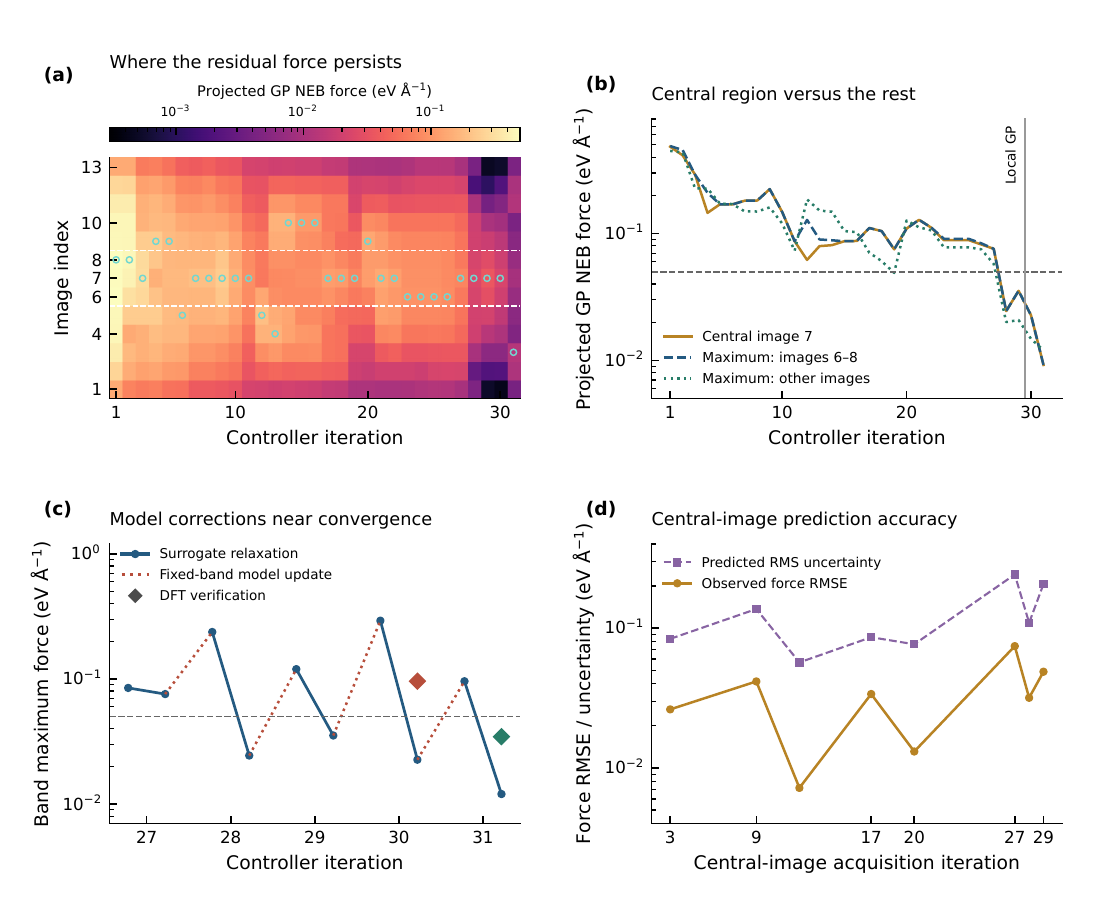}
    \caption{Evolution of projected NEB forces and GP prediction accuracy during the 96--15--GL
    calculation.
    (a) Distribution of GP-predicted projected NEB forces across the 13 moving images as a function
    of controller iteration.
    (b) Evolution of the projected GP NEB force at the central image (image 7), the maximum
    force among images 6--8, and the maximum force among the remaining images. The vertical line marks activation of the local GP.
    (c) Evolution of the maximum projected band force during the final optimization stages,
    distinguishing surrogate relaxation, fixed-band model updates, and independent
    full-band DFT verification. The two DFT verifications yielded maximum projected forces
    of 0.0960 and 0.0346~eV/\AA{}, respectively.
    (d) Comparison of observed GP force-prediction RMSE and predicted RMS force uncertainty at
    acquired central-image configurations. Horizontal dashed lines in (b) and (c) indicate the DFT NEB force acceptance threshold of 0.050~eV/\AA{}.}
    \label{fig:central_force}
\end{figure*}

A more detailed analysis of the force evolution and prediction accuracy is presented in Fig.~\ref{fig:central_force}. The distribution of projected GP NEB forces across the moving images [Fig.~\ref{fig:central_force}(a)] demonstrates that residual forces are not confined to the central image. The comparison between the central region and the remaining images [Fig.~\ref{fig:central_force}(b)] further illustrates the need to monitor convergence across the complete band, even when a local GP is introduced near the charge-transfer region.

The final optimization stages [Fig.~\ref{fig:central_force}(c)] exhibit changes in the predicted maximum band force following model updates, reflecting the incorporation of additional first-principles
information into the surrogate. The initial DFT verification reveals residual forces
above the prescribed threshold, despite the smaller surrogate forces. Subsequent refinement and verification establish convergence of the DFT NEB band.

For the acquired central-image configurations, the predicted RMS force uncertainties generally exceed the observed force RMSE values [Fig.~\ref{fig:central_force}(d)]. This indicates that the model uncertainties provide a useful diagnostic of prediction accuracy for the sampled configurations,
although it does not establish statistical calibration throughout the entire path. Together, these results emphasize the importance of force monitoring, adaptive model updates, and independent first-principles verification.

The corresponding DFT sampling history and GP prediction-error analysis for the 96-atom ($2\times2\times4$) supercell, using the 15-image discretization, are provided in Fig.~S2 of the Supplementary Information.

\subsection{324-atom system: $[001]$ and $[111]$ migration pathways}\label{sec:results_324}

Increasing the supercell from $2 \times 2 \times 4$ to $3 \times 3 \times 6$ changes not only the cost of the underlying DFT calculation but also the dimensionality of the joint energy-force GP. Because all atoms were active in the production calculations, each retained 96-atom geometry contributes 289 scalar observations (one energy and 288 force components), whereas each 324-atom geometry contributes 973 observations. With the initial 11-geometry training band, the corresponding covariance dimensions are therefore 3179 and 10703.

The increase is substantial because dense GP factorization scales cubically with the number of observations and storage scales quadratically. The 324-atom calculation therefore represents more than a straightforward application of the same algorithm to a more expensive DFT system: it directly tests whether the energy-force surrogate and active-learning controller remain usable when each additional reference geometry introduces almost one thousand correlated observations.

The 324-atom $[001]$ workflow reached a DFT-verified band. Its final prediction audit gives a mean force RMSE of 0.0042~eV/\AA{} and mean predicted RMS force uncertainty of 0.0147~eV/\AA{}. The mean uncertainty exceeds the mean error at this verification stage; this aggregate comparison should not be interpreted as proof of componentwise coverage or calibration throughout the trajectory. The corresponding per-image GP force history, central-image force response, and DFT verification diagnostics are shown in Fig.~S7 of the Supplementary Information. The full-band DFT force is the acceptance criterion; small surrogate forces alone do not establish convergence.

\begin{figure*}[!t]
    \centering
    \includegraphics[width=0.80\textwidth]
    {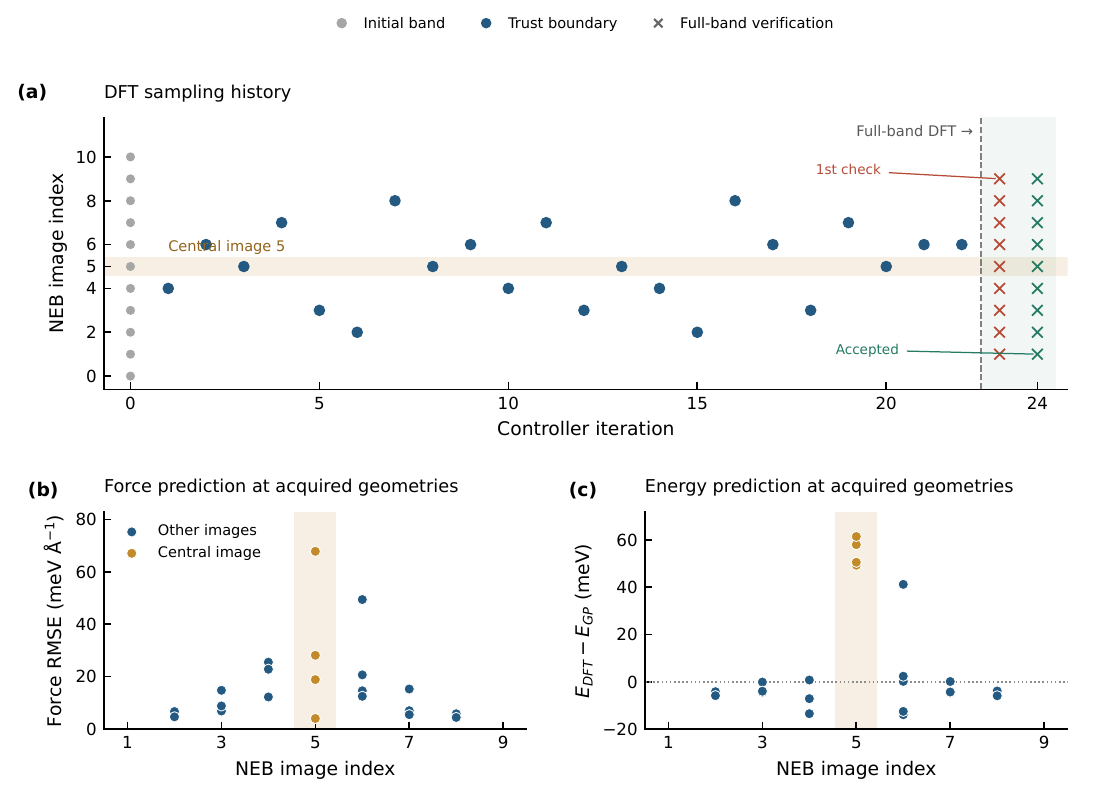}
    \caption{DFT sampling history and GP prediction 
    errors for the 324-atom, 11-image GP-assisted
    NEB calculation.
    (a) Distribution of DFT evaluations across
    NEB images and controller iterations.
    Grey markers indicate the initial band,
    blue markers indicate configurations acquired
    at trust boundaries, and crosses indicate
    full-band DFT verification.
    The first verification was unsuccessful,
    whereas the second satisfied the DFT NEB
    force convergence criterion.
    (b) Root-mean-square error (RMSE) of the
    GP-predicted Cartesian forces relative to DFT
    at acquired configurations.
    (c) Energy prediction error,
    $E_{\mathrm{DFT}}-E_{\mathrm{GP}}$,
    at the corresponding configurations.
    The central image (index 5) is highlighted
    to distinguish the charge-transfer region
    from the remaining NEB images.}
    \label{fig:dft_sampling_336}
\end{figure*}

The 96--11--G barrier is 0.1645~eV and the 324--11--G barrier is 0.1302~eV, a reduction of 0.0343 eV (20.9\%). Both bands satisfy the same DFT NEB force threshold. The decrease is consistent with cell-size sensitivity, but the $2\times2\times2$ and $\Gamma$-only Brillouin-zone samplings differ. Without matched $k$-point convergence tests, the full difference cannot be assigned uniquely to finite-size effects. Supercell-size and image-interaction effects are well-recognized considerations in charged-defect electronic-structure calculations \cite{Freysoldt2014}, although the present hopping-barrier difference should not be identified with a specific defect finite-size correction. A force threshold also does not independently bound the remaining barrier error.
A difference of a few tens of meV matters for transport because hopping rates depend exponentially on the activation free energy. The 96-atom system remains useful for algorithm development and reproducibility tests, while the larger cell provides a distinct physical reference. Agreement across the smaller-cell numerical setups does not remove finite-cell effects. The two cell sizes neither establish the infinite-cell limit nor isolate the contribution of each periodic dimension or the altered $k$-point sampling. The larger cell therefore serves as the present large-cell reference rather than a demonstrated infinite-size limit.

An additional 324-atom, 13-image $[001]$ calculation targeting the Ti157--Ti163 endpoints did not satisfy the 0.050~eV/\AA{} DFT force criterion within 75 controller iterations. Its 17 full-band verification attempts and additional acquisitions required 258 geometry-level DFT evaluations. Even the closest verification to the force threshold yielded 0.05149~eV/\AA{}, so no converged barrier is assigned. This attempt demonstrates that repeated GP refinement does not guarantee DFT acceptance within a fixed computational budget. The complete non-monotonic verification history is documented in the Supplementary Information; it is not used to assess image-count convergence.

To test the workflow on a crystallographically distinct hop, a 13-image GP-assisted NEB calculation was performed in the 324-atom ($3\times3\times6$) rutile supercell between Ti157 and Ti302. The prepared spin-initialization inputs set \texttt{DM.InitSpin} on Ti157 in polaron A and on Ti302 in polaron B, in contrast to the Ti157--Ti163 $[001]$ endpoints. Their separation is 3.573~\AA{} in the supplied endpoint geometry, and the displacement $(a/2,a/2,c/2)$ specifies the $[111]$ direction. These inputs define the intended localized-polaron endpoints; they do not alone verify the self-consistent electronic state at every intermediate image.

The $[111]$ calculations employed a global Mat\'ern-$5/2$ Gaussian process with $\sigma_f=0.225$~eV and $\ell=0.0300$~\AA{}, incorporating all 324 atoms and retaining up to 40 training geometries. No local central GP was activated for these calculations. The resulting migration pathways were accepted only after full-band DFT verification against the prescribed projected NEB force criterion of 0.050~eV/\AA{}.

The global-GP controller acquired eight additional DFT configurations during optimization of the $[111]$ band. The selected images were those reaching a geometric trust boundary, rather than necessarily those with the largest predicted GP uncertainty. Although these acquisitions lay in the first half of the band, all moving images were evaluated at every full-band DFT verification. This distinction is documented in Fig.~\ref{fig:dft_sampling_111}, with the detailed trust-boundary and uncertainty comparison in Fig.~S5 of the Supplementary Information.

\begin{figure*}[!t]
\centering
\includegraphics[width=0.95\textwidth]{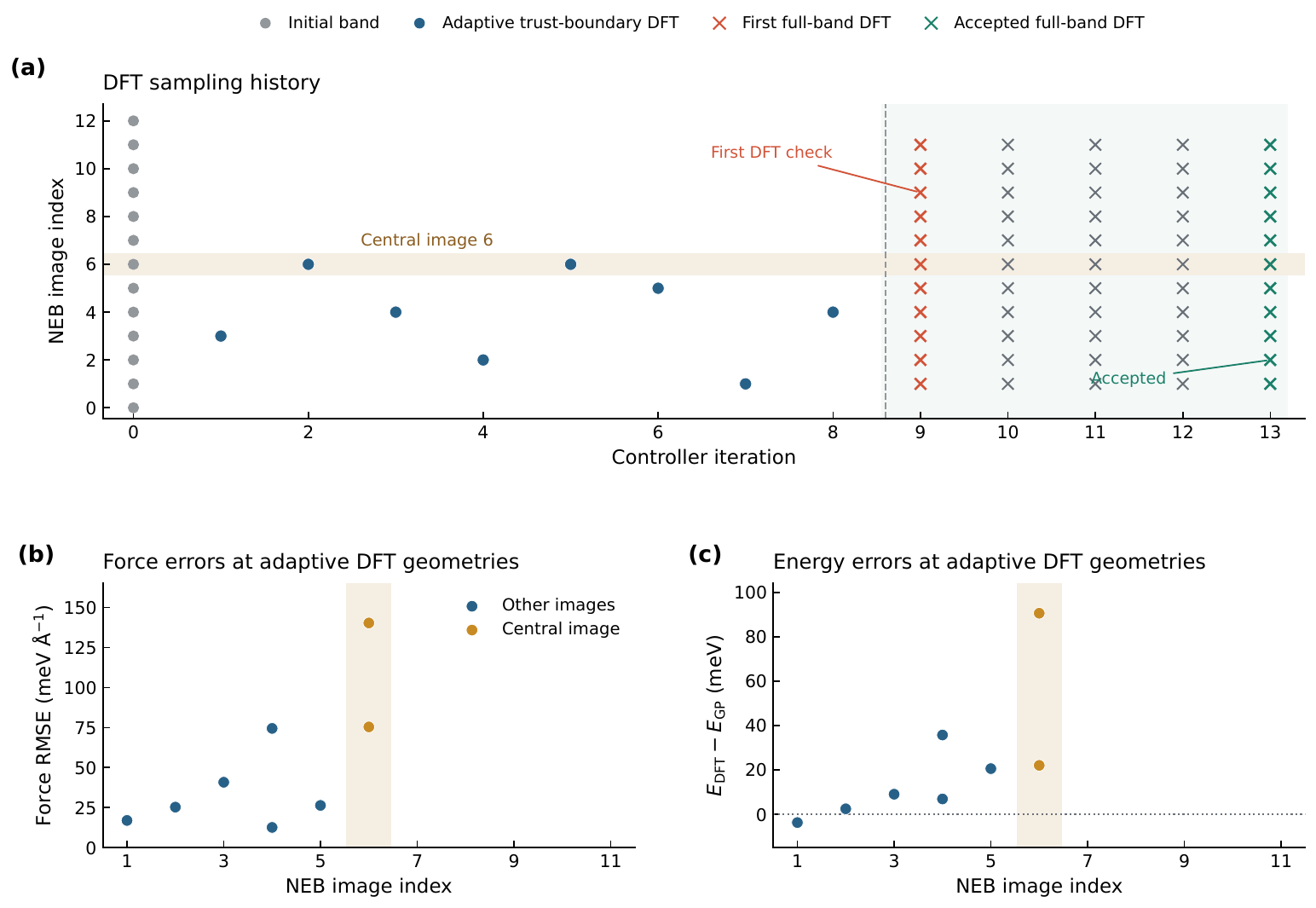}
\caption{DFT sampling and GP prediction errors for the 324-atom, 13-image $[111]$ polaron hop (run A). (a) Grey points show the initial 13-image DFT band, blue points mark eight additional acquisitions at geometric trust boundaries, and columns of crosses show five full-band verifications (11 moving images in each). The first verification failed the DFT force criterion, whereas the fifth, highlighted in green, was accepted; intermediate checks are grey. (b) Cartesian force RMSE and (c) signed energy error $E_{\mathrm{DFT}}-E_{\mathrm{GP}}$ at the eight \emph{adaptive} acquisition geometries. Central image 6 is highlighted. The lack of blue points at images 7--11 denotes the acquisition history only: those images were included in the initial DFT band and all five verifications.}
\label{fig:dft_sampling_111}
\end{figure*}

The first full-band DFT verification gave a barrier of 0.1873~eV but a maximum projected DFT NEB force of 0.3844~eV/\AA{}, far above the 0.050~eV/\AA{} threshold. Four further verification cycles progressively reduced the DFT force residual; the fifth yielded a \emph{DFT-verified} barrier of 0.1826~eV and maximum projected DFT NEB force of 0.04469~eV/\AA{}. The final global-GP band barrier prior to DFT evaluation was 0.1883~eV, showing that a small surrogate force residual is not interchangeable with first-principles barrier and force verification. The complete five-stage verification history is given in Table~S1 of the Supplementary Information. Figure~S8 of the Supplementary Information provides the complementary $[111]$ central-force study, including image-resolved GP forces, the response of central image~6, and the comparison between surrogate-force evolution and DFT-verification forces. Together, Figs.~S7 and S8 document the different convergence histories of the two accepted 324-atom hopping directions without implying that a local central GP was activated.

This accepted $[111]$ run required 13 initial DFT geometries, eight adaptive acquisitions and five verification blocks of 11 moving images, totaling 76 geometry-level DFT evaluations (Table~\ref{tab:dft_accounting}). The final full-band prediction audit yielded a mean force RMSE of 0.00190~eV/\AA{} and mean predicted RMS force uncertainty of 0.00292~eV/\AA{}; the aggregate error-to-uncertainty ratio was 0.650. Earlier central-image acquisitions showed substantially larger energy-prediction errors, so this final audit does not establish calibrated uncertainty throughout optimization or electronic-branch continuity.
A separate calculation with the same nominal Ti157--Ti302 $[111]$ endpoints and the same controller parameters reproduced the eight-image adaptive acquisition sequence and converged after five verification passes. Its DFT-verified barrier was 0.1828~eV and its maximum projected DFT NEB force was 0.04624~eV/\AA{}, compared with 0.1826~eV and 0.04469~eV/\AA{} in the first run. Both required 76 geometry-level DFT evaluations. The difference between the two accepted barriers is approximately 0.2~meV. As the two runs have matching acquisition sequences and closely matched inputs, this is a numerical repeatability check, not a statistical uncertainty estimate over independent hopping trajectories.

Compared with the accepted 324-atom, 11-image $[001]$ band (0.1302~eV), the two accepted $[111]$ bands are higher by approximately 52.4 and 52.6~meV, respectively. The distinct endpoint displacements are consistent with direction-dependent polaron migration energetics. Because the two pathways were optimized with different image counts and GP settings, the comparison establishes the difference between these accepted pathways rather than an isolated dependence on crystallographic direction. No matched conventional NEB baseline is available at 324 atoms.

The same criterion was applied to both 324-atom pathways. The $[001]$ band yielded a barrier of 0.1302~eV and a maximum projected DFT NEB force of 0.0445~eV/\AA{}, while the $[111]$ band yielded 0.1826~eV and 0.0447~eV/\AA{}, respectively. Consequently, the reported barriers are not surrogate predictions: they are energies of explicitly DFT-evaluated bands satisfying the chosen first-principles NEB force criterion.

Table~\ref{tab:dft_accounting} summarizes every geometry-level DFT evaluation, including verification attempts and the unsuccessful 324-atom $[001]$, 13-image calculation. For that unsuccessful attempt, the 258 evaluations are a recorded computational cost without an accepted barrier; they are not an acceleration measurement against conventional 324-atom NEB. Only the 96-atom, 13-image $[001]$ system has a matched conventional-NEB baseline. Multi-stage SCF solves of the same geometry are counted once.

\begin{table*}[!t]
\centering
\caption{DFT evaluation counts and migration barriers for the GP-assisted NEB calculations. Evaluations are grouped into initial-band calculations, additional DFT calculations during optimization, and full-band DFT verification.  The dagger indicates a calculation that did not satisfy the DFT NEB force convergence criterion of 0.050~eV/\AA{}; no converged barrier is reported. }
\label{tab:dft_accounting}
\begin{tabular}{lccccc}
\hline
Label & Initial band & Optimization & Full-band verification & Total DFT & Barrier (eV) \\
\hline
96--11--G  & 11 & 14 & 18 & 43 & 0.1645 \\
96--13--G  & 13 & 23 & 11 & 47 & 0.1637 \\
96--15--GL & 15 & 34 & 26 & 75 & 0.1640 \\
324--11--G & 11 & 22 & 18 & 51 & 0.1302 \\
324--13--G$_{[001]}^{\dagger}$ & 13 & 58 & 187 & 258 & --- \\
324--13--G$_{[111]}$ (A) & 13 & 8 & 55 & 76 & 0.1826 \\
324--13--G$_{[111]}$ (B) & 13 & 8 & 55 & 76 & 0.1828 \\
\hline
\end{tabular}
\end{table*}

\begin{figure*}[!t]
    \centering
    \includegraphics[width=0.80\textwidth]{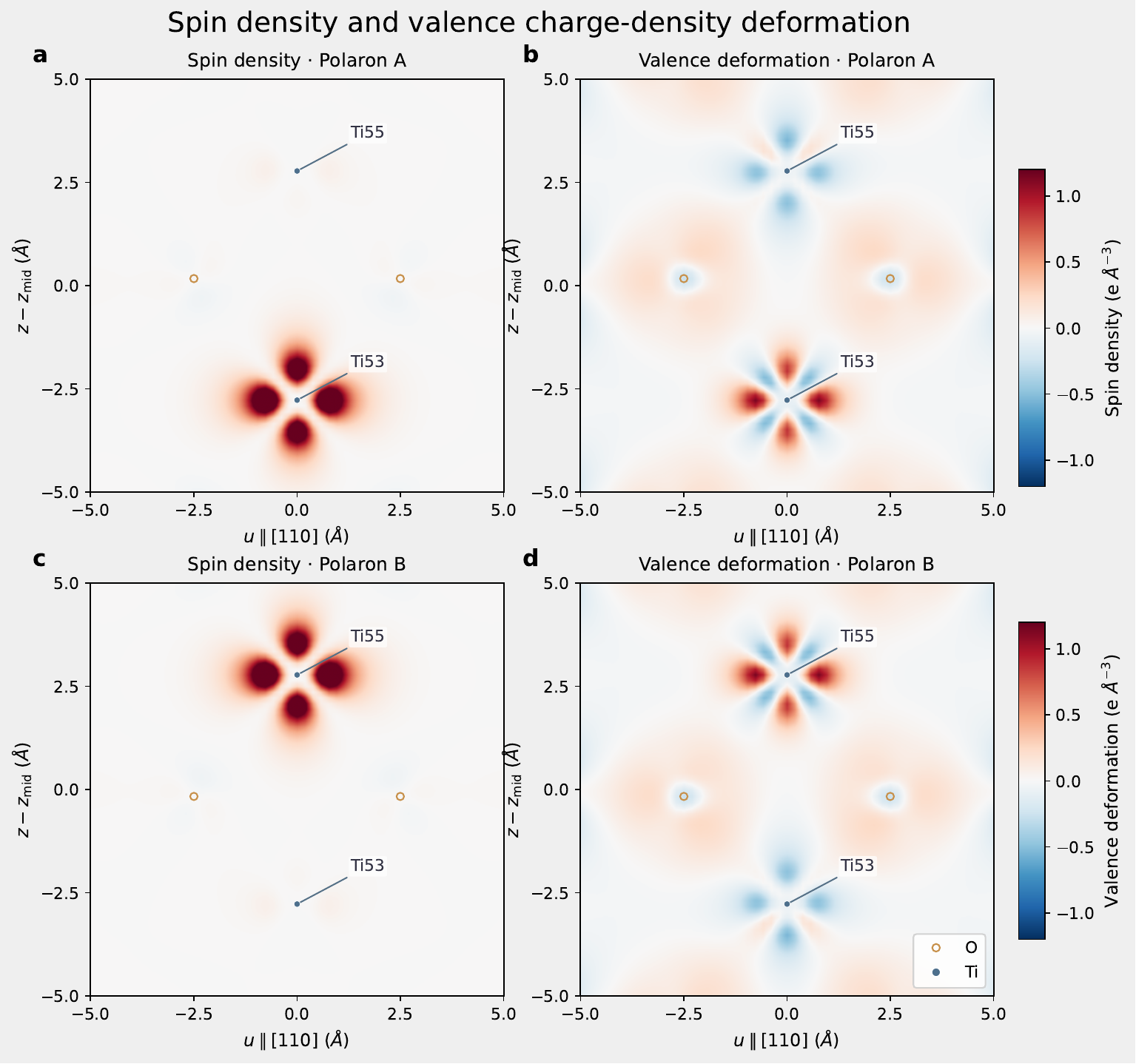}
    \caption{Comparison of spin density (left) and valence charge-density deformation (right) along $[001]$ for localized electron polarons at Ti53 (A, upper row) and Ti55 (B, lower row). All panels show the same $[-5,5]$~\AA{} longitudinal window.  Density scales are in $e$/\AA$^3$ and are shared between endpoints within each column.}
    \label{fig:spin_charge_combined}
\end{figure*}

The 324-atom $[001]$ barrier of 0.1302~eV is higher than the 58~meV barrier reported by Morita et al. for a $3\\times3\\times6$ rutile supercell using HSE06 \cite{Morita2023}. Although the cell dimensions are comparable, the electronic-structure treatments differ: the present calculations use PBE+$U$ and localized numerical atomic orbitals, whereas Morita et al. used a screened hybrid functional and plane waves. Differences in the treatment of localized charge and structural relaxation can affect the resulting adiabatic energy profile \cite{Elmaslmane2018,Reticcioli2022}. The comparison is therefore a reference to the sensitivity of polaron energetics to the electronic-structure model, not a test of GP interpolation accuracy.

In the structural analysis associated with the localized Ti53 state illustrated here, the mean nearest-neighbor Ti--O distance increases from 1.9611~\\AA{} in the pristine reference to 2.0300~\\AA{} in the relaxed polaron geometry (an average increase of 0.0689~\\AA{}). The four shorter and two longer bonds in the reference shell are consistent with the inequivalent Ti--O environments of rutile \cite{Baur1956,Szwacki2023}. The mean Ti--O distance around the site labelled Ti55 remains approximately 1.9607~\\AA{}. This preferential local expansion, together with the displayed spin-density localization, illustrates the structural response associated with the excess electron.

The endpoint-resolved spin density and valence charge-density deformation are compared in Fig.~\ref{fig:spin_charge_combined}. The longitudinal sections in the $[110]$--$[001]$ plane show that the spin polarization is predominantly localized around Ti53 in configuration A [Fig.~\ref{fig:spin_charge_combined}(a)] and Ti55 in configuration B [Fig.~\ref{fig:spin_charge_combined}(c)]. The localized distributions exhibit a lobed spatial character consistent with the Ti $3d$ contribution to the polaron state. The corresponding valence charge-density deformation maps [Figs.~\ref{fig:spin_charge_combined}(b,d)] show pronounced regions of positive and negative deformation surrounding the respective Ti sites.

The total valence charge-density deformation is defined as
\begin{equation}
\Delta\rho_{\mathrm{tot}}(\mathbf{r}) = \rho_{\mathrm{SCF}}(\mathbf{r}) - \rho_{\mathrm{atm}}(\mathbf{r}),
\label{eq:charge_redistribution}
\end{equation}
where $\rho_{\mathrm{SCF}}$ is the spin-summed self-consistent valence electron density and $\rho_{\mathrm{atm}}$ is the superposition of atomic reference valence densities. Positive and negative values indicate electron-density accumulation and depletion relative to this reference, respectively. Unlike the spin density, which characterizes the spatial distribution of spin polarization, the deformation density
also includes contributions associated with chemical bonding and should not be interpreted as the isolated excess-electron density.

Together, the spatial localization of the spin density, the deformation-density patterns, and the previously quantified Ti--O bond-length changes provide complementary evidence of the electronic and structural response associated with the localized small-polaron endpoints. The corresponding transverse sections through
Ti53 and Ti55 are provided in Fig.~S3 of the Supplementary Information.

The methodological contribution of this work is the application and evaluation of trust-limited global GP optimization, optional central-region GP refinement, and explicit DFT acceptance in a localization-sensitive, periodic polaron system. GP-assisted path optimization and first-principles force verification are established approaches \cite{Koistinen2017,Koistinen2019,Torres2019,Onyango2025}; the present calculations establish their operation for small-polaron hopping across the tested image counts and directions. The unsuccessful 324-atom $[001]$ case further illustrates that the controller does not equate a stable candidate barrier with an accepted DFT pathway.

The limitations concern the electronic-structure description, possible competing localized electronic solutions, and the cost of dense derivative GP regression as the number of atoms and retained geometries increases. DFT force verification establishes convergence of the optimized nuclear band for the electronic solutions obtained, but does not by itself establish continuous occupation of the same electronic branch. The present numerical comparisons also do not isolate the contributions of individual controller components or determine finite-temperature hopping rates. These limitations do not affect the documented DFT force-based acceptance of the converged bands.

\FloatBarrier

\section{Conclusions and Outlook}
A trust-limited energy--force GP was applied to NEB optimization of electron small-polaron hopping in rutile TiO$_2$. Geometric trust boundaries trigger targeted DFT sampling, and independent full-band DFT evaluations determine final convergence. A separately calibrated central-region GP was activated in the 96-atom, 15-image calculation while the global surrogate remained responsible for the remainder of the band.

Three DFT-accepted 96-atom $[001]$ calculations yield barriers of 0.1645, 0.1637, and 0.1640~eV. At 324 atoms, the accepted $[001]$ barrier is 0.1302~eV and two closely matched $[111]$ calculations give 0.1826 and 0.1828~eV. A separate 324-atom, 13-image $[001]$ attempt was not accepted after 258 DFT evaluations, underscoring the need to distinguish candidate barriers from DFT force-converged results.

For the matched 96-atom, 13-image $[001]$ benchmark, GP assistance reduces geometry-level DFT evaluations from 255 to 47, an 81.6\% reduction, while reproducing the conventional NEB barrier within approximately 0.1~meV. This demonstrates a practical reduction in first-principles sampling for the tested polaron migration pathway while retaining explicit DFT verification of its NEB forces.

\section*{Acknowledgements}
This research by A. K. S. was conducted during a Secondment at Simune Atomistics and was supported by Horizon Europe MSCA Postdoctoral Fellowship grant no. 101153119. 
I. V. L. acknowledges funding from the Diputaci\'on Foral de Gipuzkoa under the Quantum Gipuzkoa program and the technical and human support provided by the DIPC Supercomputing Center. 

\section*{Supporting Information}
Additional mathematical details of the joint energy--force Gaussian-process formulation, training-data management, and DFT evaluation accounting, together with supplementary atomic-structure, sampling-history, electronic-density, and 324-atom $[001]$ and $[111]$ central-force analysis figures (PDF).

\section*{Data Availability Statement}

The data supporting the findings of this study, including the output logs from the first-principles calculations and NEB optimizations, as well as the source code developed for the GP-assisted NEB workflow, are available from the corresponding authors upon reasonable request.

\printbibliography

\end{document}


\maketitle
\vspace{-0.8em}
\begin{center}
\rule{0.84\linewidth}{0.4pt}
\end{center}
\vspace{0.3em}

This document provides the covariance expressions, linear-system formulation,
posterior predictions, uncertainty diagnostics, and training-data update equations
for the joint energy--force Gaussian-process (GP) description in the main text.
The derivations state their metric, differentiability, and numerical assumptions
without implying that every optional solver or calibration method was used in
production. Supplementary figures illustrate the periodic supercell, the 96-atom
DFT sampling history, transverse electronic-density sections, and force-evolution
diagnostics for the DFT-accepted 324-atom migration pathways.

\section{Joint energy--force Gaussian-process formulation}
\label{sec:sup_gp}

\subsection{Coordinate metric and Mat\'ern covariance}
\label{sec:sup_kernel}

Let $N_a$ be the number of active atoms, $D=3N_a$ the number of active Cartesian
components, and $N_g$ the number of retained first-principles training geometries.
For Cartesian coordinates $\mathbf q,\mathbf q'\in\mathbb R^D$, define
$\Delta_j=q_j-q'_j$ for Cartesian component $j$ using a locally consistent minimum-image mapping. The
root-mean-square (RMS) distance metric and the Mat\'ern-$5/2$ energy covariance are
\begin{equation}
\begin{aligned}
\rho^2(\mathbf q,\mathbf q')&=\frac{1}{N_a}\sum_{j=1}^{D}\Delta_j^2,
&u&=\frac{\sqrt 5\,\rho}{\ell},\\[0.3em]
k(\mathbf q,\mathbf q')&=\sigma_f^2\left(1+u+\frac{u^2}{3}\right)e^{-u}.
\end{aligned}
\label{eq:sup_rho_and_k}
\end{equation}
Here $\ell$ is the correlation length and $\sigma_f$ has units of energy. The
metric is normalized by the number of atoms rather than by $D$.
The derivative formulas below assume that the periodic-image mapping remains
fixed locally while derivatives are evaluated \cite{Rasmussen2006,Koistinen2017}.

\subsection{Energy and force covariance blocks}
\label{sec:sup_covariance}

With Cartesian-component indices $a,b\in\{1,\ldots,D\}$ and the force convention $F_a=-\partial E/\partial q_a$, the covariance blocks
follow by differentiating the energy kernel \cite{Rasmussen2006,Koistinen2017}:
\begin{equation}
\begin{aligned}
K_{\mathrm{EE}}&=k,
&K_{\mathrm{EF}_b}&=-\partial_{q'_b}k,\\[0.4em]
K_{\mathrm{F}_a\mathrm{E}}&=-\partial_{q_a}k,
&K_{\mathrm{F}_a\mathrm{F}_b}&=\partial_{q_a}\partial_{q'_b}k.
\end{aligned}
\label{eq:sup_covariance_blocks}
\end{equation}
For the RMS Mat\'ern metric in Eq.~\eqref{eq:sup_rho_and_k}, define
$c=5\sigma_f^2/(3\ell^2)$. The resulting expressions are
\begin{equation}
\begin{aligned}
K_{\mathrm{EF}_b}
  &=-\frac{c}{N_a}(1+u)e^{-u}\Delta_b,\\[0.4em]
K_{\mathrm{F}_a\mathrm{E}}
  &=\frac{c}{N_a}(1+u)e^{-u}\Delta_a,\\[0.4em]
K_{\mathrm{F}_a\mathrm{F}_b}
  &=\frac{c}{N_a}(1+u)e^{-u}\delta_{ab}
    -\frac{5c}{N_a^2\ell^2}e^{-u}\Delta_a\Delta_b.
\end{aligned}
\label{eq:sup_covariance_derivatives}
\end{equation}
These expressions remain finite at coincident configurations. In particular,
the prior variance of a Cartesian force component is
$5\sigma_f^2/(3N_a\ell^2)$. The logged quantity
$g=\sigma_f/(\ell\sqrt{N_a})$ is therefore a characteristic force scale, not
by itself evidence that the force observations are independently reweighted.
The component prior standard deviation is $\sqrt{5/3}\,g$.

Stack all energy observations first and all active Cartesian force components
second; any consistently applied permutation defines the same model:
\begin{equation}
\begin{aligned}
\mathbf y&=\bigl(E_1,\ldots,E_{N_g},
 \mathbf F_{1,\mathrm A}^{\mathsf T},\ldots,
 \mathbf F_{N_g,\mathrm A}^{\mathsf T}\bigr)^{\mathsf T},\\[0.3em]
\mathbf z&=\mathbf y-\mathbf m_y,\\[0.3em]
\mathbf K&=\begin{bmatrix}
\mathbf K_{\mathrm{EE}}&\mathbf K_{\mathrm{EF}}\\
\mathbf K_{\mathrm{FE}}&\mathbf K_{\mathrm{FF}}
\end{bmatrix},\\[0.3em]
\mathbf C&=\mathbf K+\boldsymbol\Lambda.
\end{aligned}
\label{eq:sup_joint_gp_vectors}
\end{equation}
The total observation dimension is $M=N_g(1+D)$. The diagonal matrix
$\boldsymbol\Lambda$ contains energy variances $\boldsymbol\eta_E^2$ and
force variances $\boldsymbol\eta_F^2$ in their respective units, together with
any compatible numerical regularization. The recorded parameter summaries do not establish the
numerical regularization or whether observation scaling was enabled, so neither
is assigned a numerical value here.
If a diagonal observation scaling $\mathbf S$ is used, it must be applied
consistently:
$\widetilde{\mathbf z}=\mathbf S\mathbf z$ and
$\widetilde{\mathbf C}=\mathbf S\mathbf C\mathbf S^{\mathsf T}$,
with matching transformations of prediction cross-covariances.

\subsubsection{Model training and hyperparameter calibration}
\label{sec:sup_training}

At fixed hyperparameters, training involves solving the regularized system.
A Cholesky factorization avoids computing a dense inverse \cite{Rasmussen2006}:
\begin{equation}
\mathbf C=\mathbf L\mathbf L^{\mathsf T},\qquad
\mathbf L\mathbf v=\mathbf z,\qquad
\mathbf L^{\mathsf T}\boldsymbol\alpha=\mathbf v.
\label{eq:sup_cholesky_solve}
\end{equation}
The standard log marginal likelihood is \cite{Rasmussen2006}
\begin{equation}
\begin{aligned}
\log p(\mathbf y\mid\mathbf X)
={}&-\tfrac12\mathbf z^{\mathsf T}\boldsymbol\alpha
-\sum_{j=1}^{M}\log L_{jj}\\
&-\tfrac M2\log(2\pi).
\end{aligned}
\label{eq:sup_log_marginal_likelihood}
\end{equation}
Equation~\eqref{eq:sup_log_marginal_likelihood} is the standard GP
log marginal likelihood, included as mathematical background rather than as an
assertion that all production models maximized it. The global parameters were
read from \texttt{gp\_calibration.json}; the available controller logs record
parameter values and calibration warnings, not the full global search objective
or validation partition. The 96--15--GL local model, in contrast, was calibrated
using three-fold force-prediction validation of eight acquired central geometries,
followed by uncertainty calibration.
Hyperparameter selection and solution of Eq.~\eqref{eq:sup_cholesky_solve}
are separate operations.

\subsection{Prediction of energy, forces, and uncertainty}
\label{sec:sup_prediction}

Let $\mathbf h_*=[E_*,\mathbf F_{*,\mathrm A}^{\mathsf T}]^{\mathsf T}$,
$\mathbf B_*=\operatorname{cov}(\mathbf y,\mathbf h_*)$,
and $\mathbf K_{**}=\operatorname{cov}(\mathbf h_*,\mathbf h_*)$.
The joint posterior distribution has mean and covariance \cite{Rasmussen2006}
\begin{equation}
\begin{aligned}
\boldsymbol\mu_*&=\mathbf m_*+\mathbf B_*^{\mathsf T}\boldsymbol\alpha,\\[0.4em]
\mathbf V_*&=\mathbf K_{**}-\mathbf B_*^{\mathsf T}
\mathbf C^{-1}\mathbf B_*.
\end{aligned}
\label{eq:sup_posterior_mean_var}
\end{equation}
Numerically, solve $\mathbf L\mathbf W_*=\mathbf B_*$ and compute
$\mathbf V_*=\mathbf K_{**}-\mathbf W_*^{\mathsf T}\mathbf W_*$.
The inverse in Eq.~\eqref{eq:sup_posterior_mean_var} is mathematical notation;
it is not an instruction to form a dense inverse.
The first component of $\boldsymbol\mu_*$ is the predicted energy, and the
remaining components are the active Cartesian force predictions. Within a
single GP model, $\boldsymbol\mu_F=-\nabla_{\mathbf q}\mu_E$.
The latent energy and force uncertainties are
\begin{equation}
\begin{aligned}
\sigma_E&=\sqrt{V_{*,\mathrm{EE}}},
&\sigma_{F,a}&=\sqrt{V_{*,\mathrm{F}_a\mathrm{F}_a}},\\[0.5em]
s_F&=\sqrt{\frac1D\sum_{a=1}^{D}\sigma_{F,a}^2}.
\end{aligned}
\label{eq:sup_latent_uncertainties}
\end{equation}
Observation noise is included only when predicting future noisy observations,
not the latent GP output. At an acquired geometry, predictions recorded before
inserting the corresponding training data are compared with DFT using
\begin{equation}
\begin{aligned}
\Delta E&=\mu_E-E_{\mathrm{DFT}},\\[0.4em]
\operatorname{RMSE}_F&=
\sqrt{\frac1D\sum_{a=1}^{D}
\bigl(\mu_{F,a}-F_{a,\mathrm{DFT}}\bigr)^2}.
\end{aligned}
\label{eq:sup_dft_errors}
\end{equation}
The energy-error sign must be reversed explicitly when comparing with plots
or logs using $E_{\mathrm{DFT}}-\mu_E$.
A componentwise uncertainty-calibration diagnostic is
\begin{equation}
z_{\mathrm{RMS}}=
\sqrt{\frac1D\sum_{a=1}^{D}
\frac{\bigl(\mu_{F,a}-F_{a,\mathrm{DFT}}\bigr)^2}{\sigma_{F,a}^2}}.
\label{eq:sup_z_rms}
\end{equation}
Any uncertainty floor and aggregation across held-out configurations must
match the implementation. An average error below an average predicted
uncertainty is only an aggregate diagnostic and does not guarantee
componentwise or per-image calibration.

\subsection{Adding newly evaluated configurations}
\label{sec:sup_update}

After DFT evaluation, the energy and active Cartesian forces at the selected
geometry are added as $1+D$ observations. For fixed mean and hyperparameters,
\begin{equation}
\begin{aligned}
\mathcal D_+&=\mathcal D\cup
\{(\mathbf q_+,E_+,\mathbf F_{+,\mathrm A})\},\\[0.4em]
\mathbf z_+&=\begin{bmatrix}\mathbf z\\\mathbf z_{\mathrm{new}}\end{bmatrix}.
\end{aligned}
\label{eq:sup_data_update}
\end{equation}
Here the updated observation vector is written in old-then-new order for
convenience; this is simply a consistent permutation of the energy-first
ordering in Eq.~\eqref{eq:sup_joint_gp_vectors}.
The enlarged covariance matrix has block form
\begin{equation}
\mathbf C_+=\begin{bmatrix}
\mathbf C&\mathbf B\\
\mathbf B^{\mathsf T}&\mathbf C_{\mathrm{new}}
\end{bmatrix}.
\label{eq:sup_covariance_update}
\end{equation}
The old-to-new block $\mathbf B$ contains derivative covariances from
Eqs.~\eqref{eq:sup_covariance_blocks} and
\eqref{eq:sup_covariance_derivatives}, while $\mathbf C_{\mathrm{new}}$
includes the new observation noise. Rebuilding and factorizing
$\mathbf C_+$ produces the updated posterior using
Eqs.~\eqref{eq:sup_cholesky_solve}--\eqref{eq:sup_latent_uncertainties}.
A mathematically equivalent block factorization, provided that previous
observations and hyperparameters are unchanged, is
\begin{equation}
\begin{aligned}
\mathbf W&=\mathbf L^{-1}\mathbf B,\\
\mathbf S_{\mathrm c}&=\mathbf C_{\mathrm{new}}
-\mathbf W^{\mathsf T}\mathbf W,\\[0.4em]
\mathbf L_+&=\begin{bmatrix}
\mathbf L&\mathbf 0\\
\mathbf W^{\mathsf T}&\operatorname{chol}(\mathbf S_{\mathrm c})
\end{bmatrix}.
\end{aligned}
\label{eq:sup_block_cholesky}
\end{equation}
Equation~\eqref{eq:sup_block_cholesky} is a mathematically equivalent
block-update formula, not evidence that the production controller used
incremental Cholesky updates. If the mean
is re-estimated, the residual vector must be recomputed. Hyperparameter
changes or pruning generally require rebuilding the covariance for the
retained set. Distinct geometries of the same NEB image are separate
training observations; an image label is not a unique training configuration.

\subsection{Interpretation and limits of the numerical formulation}
\label{sec:sup_numerical_scope}
The covariance formulas are defined for a fixed set of Cartesian coordinates
and a locally unchanged periodic-image assignment. A discontinuous change in
the minimum-image mapping is outside the scope of the smooth kernel derivatives
used here. Regularization, any rescaling of energy and force observations,
and hyperparameter tuning must be implemented consistently with the kernel
and its energy--force cross-covariances. The controller traces alone do not
identify all these low-level numerical choices; consequently, the equations
above distinguish mathematically valid alternatives from verified production
operations.

The workload reported in the main text counts each requested DFT geometry
once, irrespective of the number of SCF stages required to converge that
geometry. Training-set pruning does not undo a previously executed DFT
evaluation. The full-band verification includes DFT calculations for the
moving images at the same candidate band coordinates; previously prepared
endpoint energies are reused. This is an evaluation-count convention rather
than a measurement of computational runtime.

\section{Large-supercell verification and endpoint diagnostics}
\label{sec:sup_111}
The 324-atom, 13-image $[111]$ run used the initial and final polaron structures with spin initialization on Ti157 and Ti302, respectively (1-based SIESTA indices). In the prepared endpoint files, the Ti positions are $(6.8998,6.8998,5.9184)$ and $(9.1997,9.1997,7.3980)$~\AA{}, so their displacement is $(2.2999,2.2999,1.4796)$~\AA{} (3.573~\AA{} in total). The intended displacement is $[111]$ in the tetragonal rutile lattice. The mean nearest-six Ti--O separation around the seeded polaron site is 2.059~\AA{} in each input, compared with 1.961~\AA{} in the undistorted reference shell. This is a description of prepared input distortions, not a structural or charge-localization analysis of the self-consistent final states.

\begin{table}[!htbp]
\centering
\caption{Full-band DFT verification history for the two closely matched 324-atom, 13-image $[111]$ runs. The acceptance force threshold is 0.050~eV/\AA{}. Each verification re-evaluates 11 moving images.}
\label{tab:sup_111_verification}
\begin{tabular}{ccccc}
\toprule
& \multicolumn{2}{c}{Run A} & \multicolumn{2}{c}{Run B} \\
Check & $E_b$ (eV) & $F_{\max}$ (eV/\AA{}) & $E_b$ (eV) & $F_{\max}$ (eV/\AA{}) \\
\midrule
1 & 0.187264 & 0.384390 & 0.187899 & 0.387227 \\
2 & 0.183492 & 0.159741 & 0.183780 & 0.160913 \\
3 & 0.182873 & 0.106527 & 0.183093 & 0.110191 \\
4 & 0.182674 & 0.069181 & 0.182894 & 0.072205 \\
5 & 0.182583 & 0.044691 & 0.182775 & 0.046244 \\
\bottomrule
\end{tabular}
\end{table}

Both $[111]$ runs satisfy the DFT criterion only at their fifth verification. Each workload consists of 13 initial configurations, eight adaptively selected geometries, and 55 DFT evaluations from the five full-band verifications (76 per run). The first four failed verifications are included in the count. Run B's final barrier differs from run A by only 0.192~meV, but their identical adaptive acquisition sequence does not support a statistical error bar. For run A, the final full-band force RMSE (mean over images) is 0.001895~eV/\AA{} and the corresponding predicted RMS force uncertainty is 0.002917~eV/\AA{}. No local central predictor was activated during this trajectory. The production log records an initial global-hyperparameter prediction warning; final force convergence does not eliminate the need for separate electronic-state checks.

The eight adaptive $[111]$ DFT geometries were obtained at trust-boundary stops, not by greedily acquiring the image with maximum predicted force uncertainty. In both converged runs the adaptive sequence was $3,6,4,2,6,5,1,4$ (NEB moving-image indices); the left-side image that reached the maximum-atom threshold in each iteration was the acquired one. At those stops the largest maximum-atom training distance among right-side images 7--11 was only 0.0077--0.0129~\AA{}, versus the 0.015~\AA{} trust limit reached on the left. Image 10 was nevertheless the most uncertain image at iterations 3, 4 and 8, highlighting that the geometric trust-limiting event and the maximum-uncertainty location need not coincide (Fig.~S5). Both runs subsequently evaluated all eleven moving images at each of their five full-band DFT verification stages (Fig.~S6). The first run exhibited adaptive central-image force errors of 0.1403 and 0.07541~eV/\AA{} and energy errors $E_{\mathrm{DFT}}-E_{\mathrm{GP}}$ of $+0.09059$ and $+0.02206$~eV, respectively, at its two central adaptive acquisitions. Such pointwise discrepancies are not evidence by themselves of electronic-branch switching; they motivate DFT acceptance rather than surrogate-only convergence.

\begin{figure}[!htbp]
\centering
\includegraphics[width=0.98\linewidth]{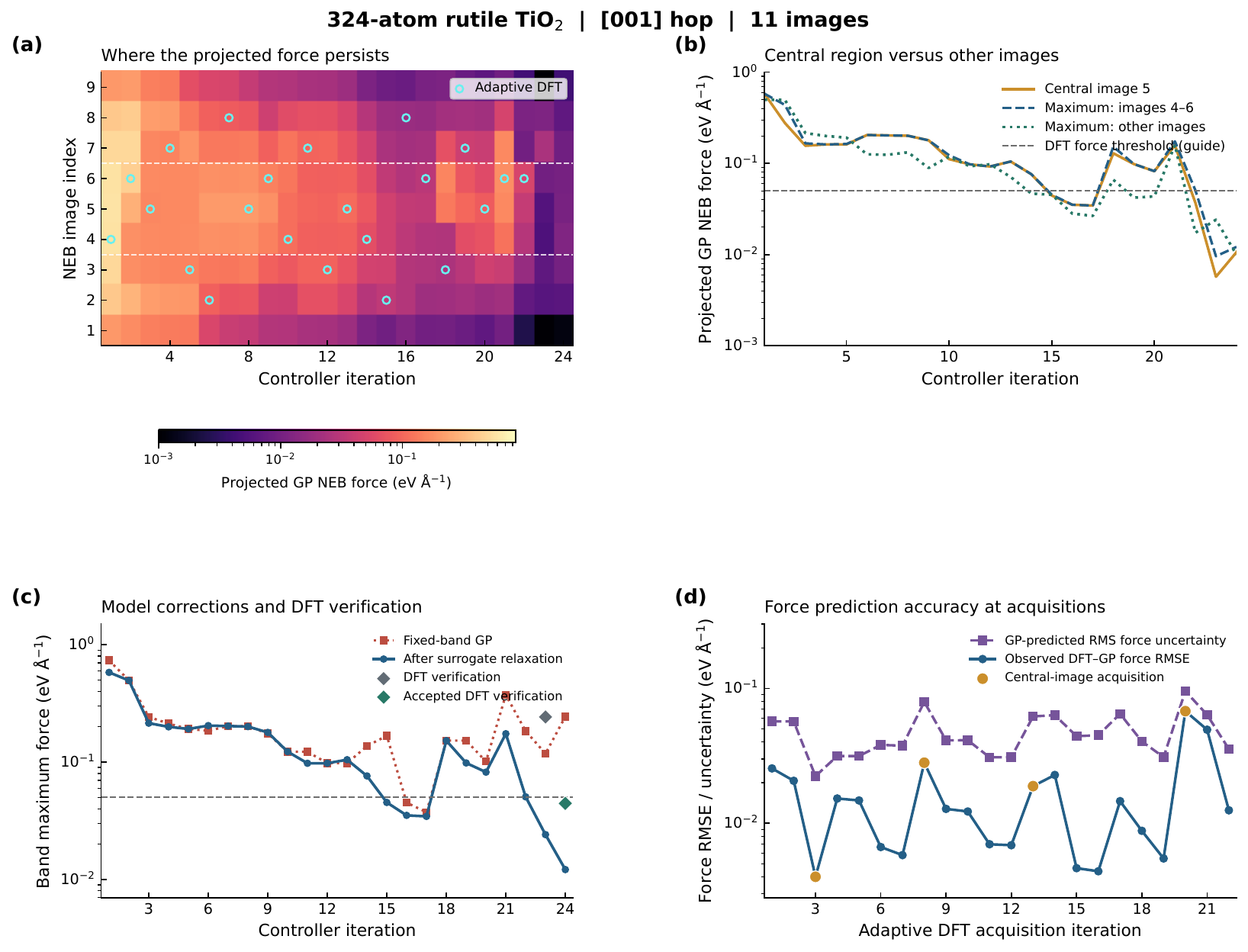}
\caption{Force evolution and surrogate prediction accuracy for the \emph{accepted} 324-atom, 11-image $[001]$ hop (Ti157--Ti163). (a) Projected GP NEB force of each moving image after surrogate movement; open circles indicate adaptive DFT acquisitions. Dashed horizontal lines bracket the central three images (4--6). (b) Projected GP force at central image~5, the maximum across images 4--6, and the maximum outside that region. (c) Fixed-band and post-relaxation GP force maxima compared with full-band DFT verification maxima. The second verification satisfies the 0.050~eV/\AA{} acceptance criterion, with $F_{\max}^{\mathrm{DFT}}=0.044493$~eV/\AA{} and $E_b=0.130157$~eV. (d) Predicted RMS GP force uncertainty and observed GP--DFT force RMSE at adaptive DFT acquisitions; central-image acquisitions are highlighted. The reference force line in GP panels is not a substitute for DFT verification.}
\label{fig:sup_force_324_001}
\end{figure}

\begin{figure}[!htbp]
\centering
\includegraphics[width=0.98\linewidth]{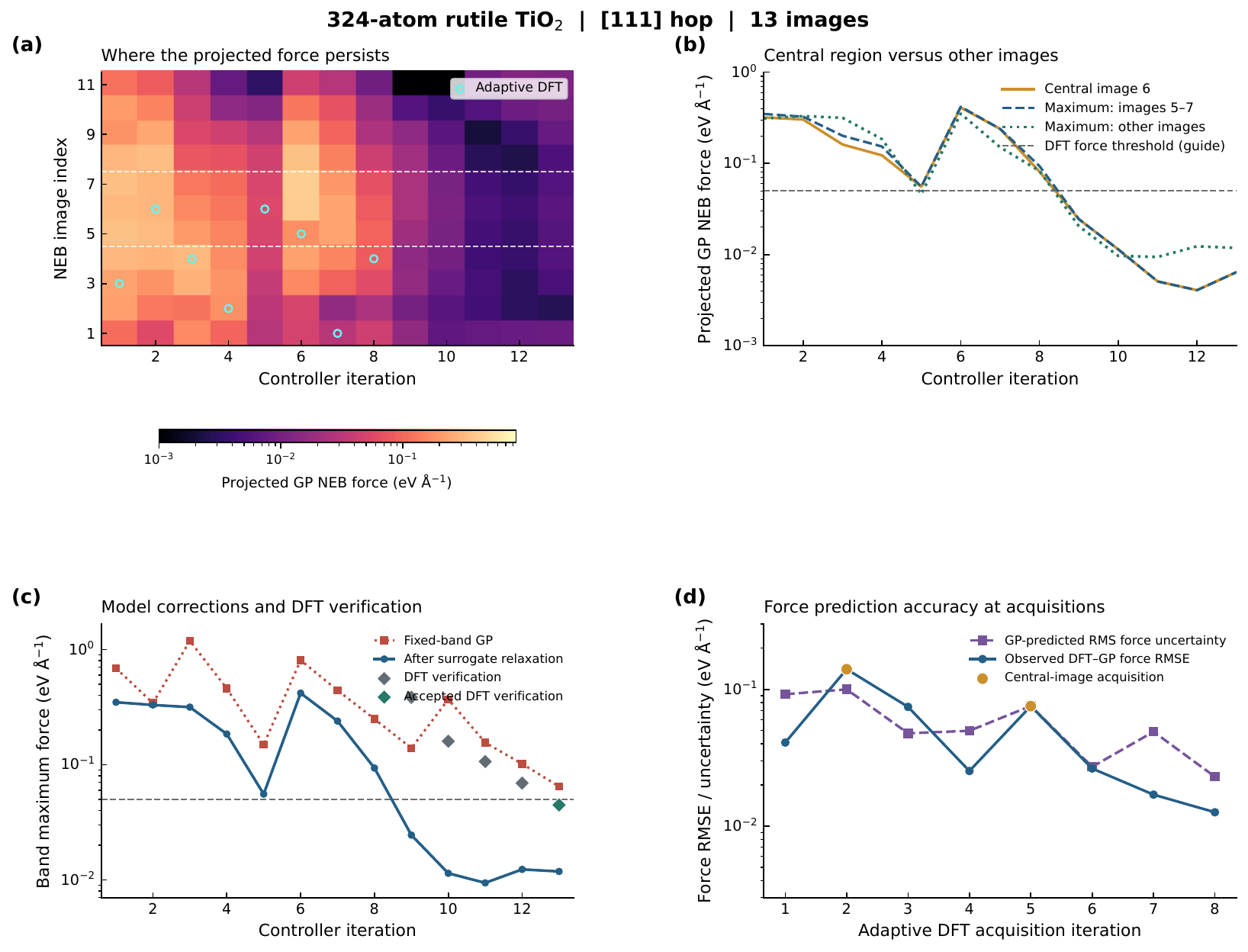}
\caption{Force evolution and surrogate prediction accuracy for the \emph{accepted} 324-atom, 13-image $[111]$ hop (Ti157--Ti302, run~A). Panels show (a) post-movement per-image projected GP NEB force with eight adaptive DFT acquisitions (open circles), (b) central image~6 and neighboring-image GP forces compared with other moving images, (c) fixed-band and post-relaxation GP force maxima together with all five full-band DFT verification maxima, and (d) GP-predicted RMS force uncertainty versus observed GP--DFT force RMSE at adaptively acquired geometries; central-image acquisitions are highlighted. The dashed lines in (a) bracket images 5--7. The fifth full-band verification satisfies the 0.050~eV/\AA{} DFT force criterion ($F_{\max}^{\mathrm{DFT}}=0.044691$~eV/\AA{}, $E_b=0.182583$~eV). The adaptive sampling markers do not represent verification DFT evaluations: all 11 moving images were re-evaluated in each verification pass. A local central GP was not activated.}
\label{fig:sup_force_324_111}
\end{figure}

The full optimization histories of the two \emph{accepted} 324-atom paths are compared in Figs.~\ref{fig:sup_force_324_001} and \ref{fig:sup_force_324_111}. The $[001]$ hop (Ti157--Ti163) employed 11 NEB images and 24 controller iterations, with 22 adaptive DFT acquisitions and two full-band DFT verification passes (51 geometry evaluations including the initial 11-image band). Its central moving image is image~5. The $[111]$ hop (Ti157--Ti302; run~A) employed 13 NEB images and 13 controller iterations, with eight adaptive acquisitions and five full-band verifications (76 geometry evaluations including the initial 13-image band); its central moving image is image~6. The respective accepted DFT projected-force maxima are 0.044493 and 0.044691~eV/\AA{}, and the barriers are 0.130157 and 0.182583~eV. These counts are for individual DFT geometry evaluations, not SCF cycles or wall-clock time.

In each force-history figure, panel (a) gives the per-image projected \emph{GP} NEB force following surrogate movement, while open markers denote only the adaptively acquired DFT geometries. Panel (b) distinguishes the central image, its three-image neighborhood, and the remainder of the band. Panel (c) separates the fixed-band GP force, the GP force after surrogate relaxation, and the independently evaluated maximum projected \emph{DFT} force at each full-band check. Panel (d) compares the predicted RMS GP force uncertainty with the observed GP--DFT force RMSE at adaptive acquisitions; markers for acquisitions of the central image are highlighted. In the $[111]$ run, adaptive samples were restricted to images 1--6 by the geometric trust-boundary stopping rule, although all 11 moving images received DFT evaluations in every full-band verification. The apparent 0.050~eV/\AA{} reference line on GP-force panels is illustrative; acceptance is based solely on the DFT-projected force in panel (c). Neither 324-atom case is presented as evidence of a sustained local-central-GP acceleration phase.

The additional 324-atom, 13-image $[001]$ calculation used spin-initialization sites Ti157 and Ti163. It produced 17 full-band DFT verification blocks without meeting the 0.050~eV/\AA{} force criterion, and its controller stopped after 75 iterations. The output includes a transient 0.6782~eV barrier at 1.166~eV/\AA{} maximum force, indicating an anomalous energy/force excursion whose cause cannot be uniquely identified from the controller log. The candidate barrier near 0.115~eV at later stages was never accepted and is therefore not included as a converged migration barrier. The total count of 258 geometry-level DFT evaluations comprises 13 initial geometries, 58 adaptive acquisitions, and $17\times11=187$ full-band verification evaluations; the retained-training cap of 40 did not reduce historical DFT costs.

\begin{table}[!htbp]
\centering
\small
\caption{Full-band DFT verification history for the unaccepted 324-atom, 13-image $[001]$ run. No verification satisfies the 0.050~eV/\AA{} acceptance criterion; barriers are candidate estimates only.}
\label{tab:sup_00113_failure}
\begin{tabular}{ccc}
\toprule
Check & Barrier estimate (eV) & $F_{\max}$ (eV/\AA{}) \\
\midrule
1 & 0.185225 & 0.096224 \\
2 & 0.144931 & 0.072388 \\
3 & 0.134532 & 0.105038 \\
4 & 0.132247 & 0.054835 \\
5 & 0.133038 & 0.094626 \\
6 & 0.133791 & 0.103507 \\
7 & 0.114685 & 0.214605 \\
8 & 0.111802 & 0.165506 \\
9 & 0.111128 & 0.153465 \\
10 & 0.678242 & 1.166084 \\
11 & 0.111672 & 0.200480 \\
12 & 0.115766 & 0.195651 \\
13 & 0.115025 & 0.061058 \\
14 & 0.114920 & 0.051490 \\
15 & 0.114902 & 0.066358 \\
16 & 0.115210 & 0.112572 \\
17 & 0.115475 & 0.160055 \\
\bottomrule
\end{tabular}
\end{table}

\section{Supplementary figures}
\label{sec:sup_figures}

\begin{figure}[!htbp]
\centering
\includegraphics[width=0.82\linewidth]{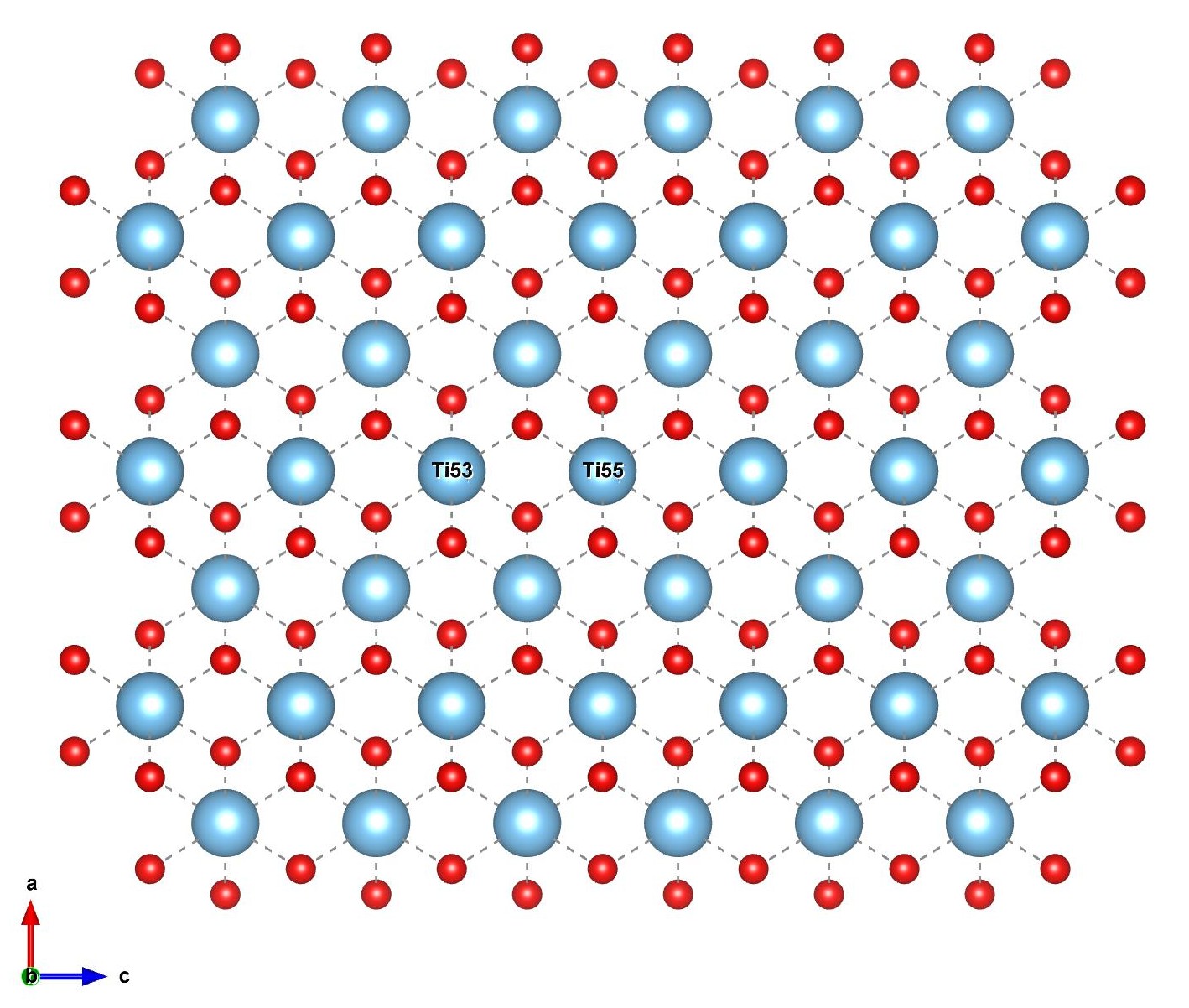}
\caption{Atomic structure of the $3\times3\times6$ rutile TiO$_2$
supercell used for the 324-atom calculation. The [001] hopping
pathway is indicated by the two labelled Ti positions (schematic labels Ti53 and Ti55, not the 324-atom SIESTA input indices). The computed [001] path uses SIESTA indices Ti157 and Ti163.}
\label{fig:sup_supercell_336}
\end{figure}

\begin{figure}[!htbp]
\centering
\includegraphics[width=0.98\linewidth]{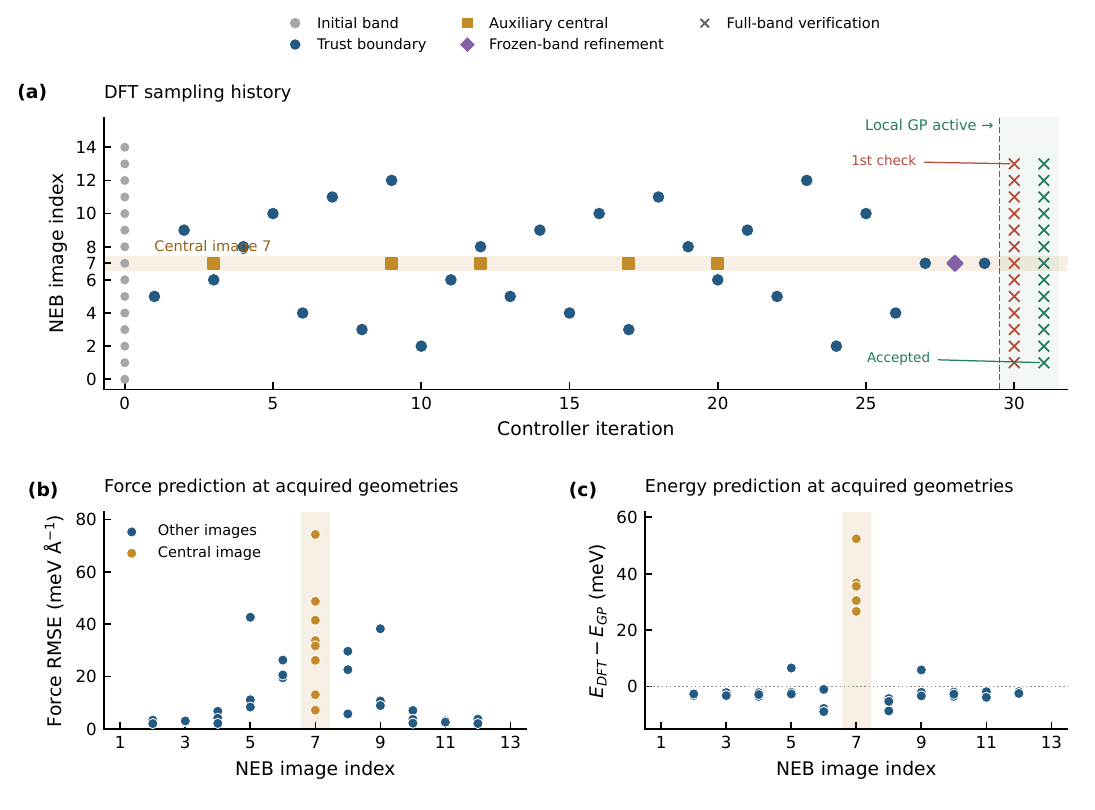}
\caption{DFT sampling history and GP prediction-error diagnostics for the
96-atom ($2\times2\times4$), 15-image GP-assisted NEB calculation.}
\label{fig:sup_sampling_224}
\end{figure}

\begin{figure}[!htbp]
\centering
\includegraphics[width=0.95\linewidth]{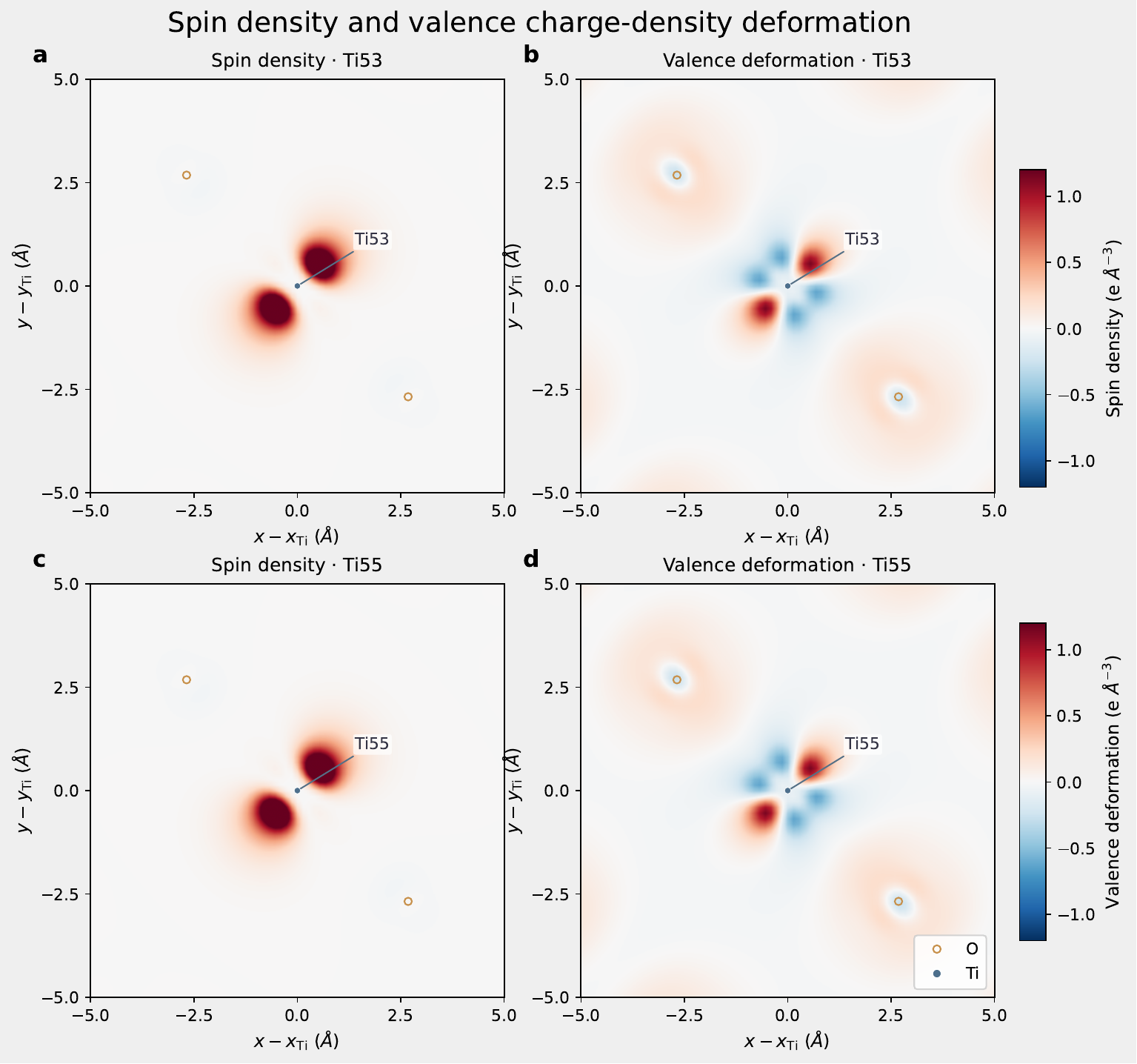}
\caption{Transverse comparison of the spin-channel difference (left column)
and valence charge-density deformation (right column) for the localized-polaron
endpoints Ti53 (upper row) and Ti55 (lower row). All spatial axes cover
$-5$ to $5$~\AA{}. Both quantities are in e/\AA$^3$. The spin-channel difference represents spin density when the atomic reference is unpolarized.}
\label{fig:sup_transverse_comparison}
\end{figure}

\begin{figure}[!htbp]
\centering
\includegraphics[width=\linewidth]{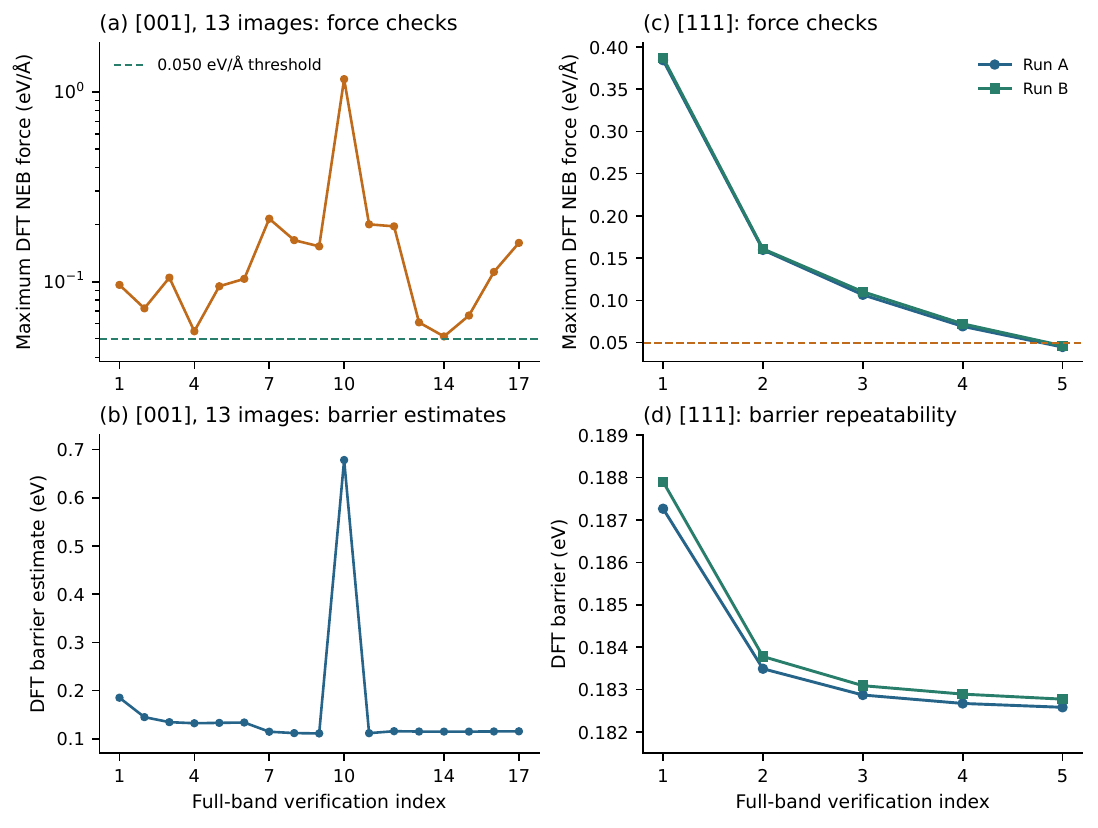}
\caption{DFT verification histories at 324 atoms. (a,b) The 13-image $[001]$ attempt, including an anomalous high-energy/high-force excursion, never satisfied the projected force threshold; the candidate barriers are not accepted results. The force axis in (a) is logarithmic. (c,d) Closely matched $[111]$ runs A and B both satisfy the threshold at their fifth verification and yield barriers differing by 0.192~meV.}
\label{fig:sup_324_verification_comparison}
\end{figure}

\begin{figure}[!htbp]
\centering
\includegraphics[width=\linewidth]{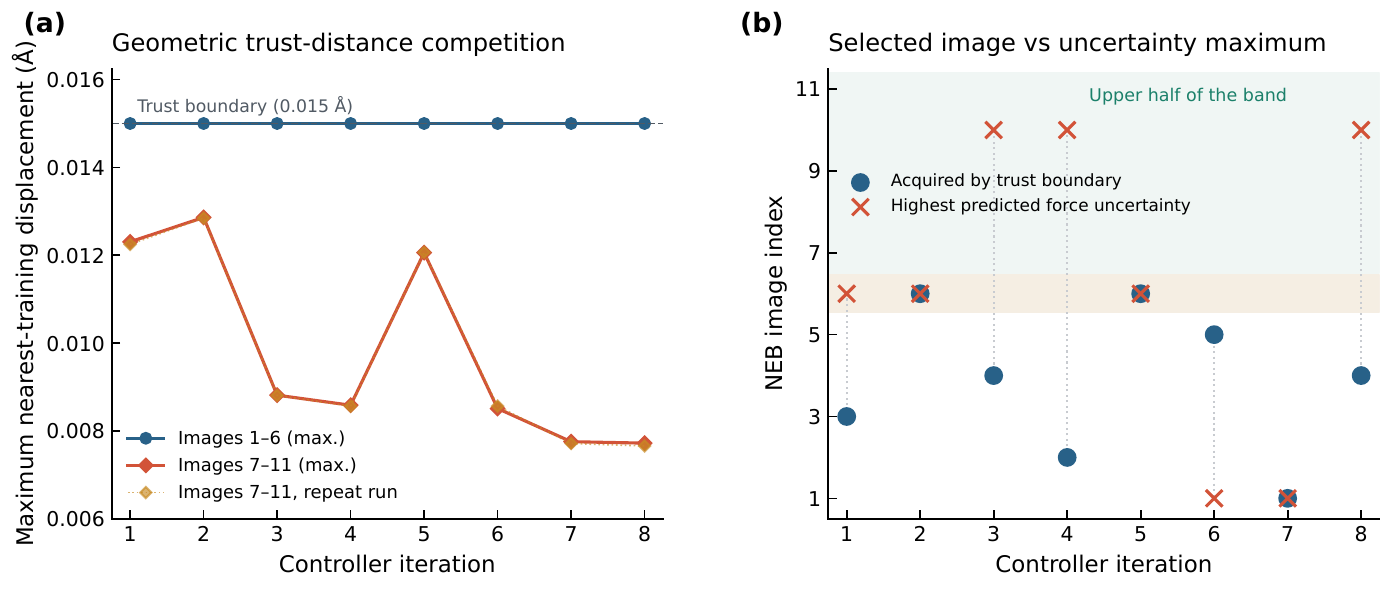}
\caption{Geometric explanation of one-sided adaptive DFT selection for the 324-atom, 13-image $[111]$ hop. (a) Largest maximum-atom displacement relative to the retained DFT training set among images 1--6 and 7--11, immediately before each trust-triggered acquisition. Left-side displacements reach the imposed 0.015~\AA{} limit, while right-side displacements remain below it. The faint right-side repeat-run curve shows the same pattern for run B. (b) Trust-boundary-acquired image indices versus indices of maximum predicted force uncertainty in run A; image 10 is most uncertain in three iterations without being selected. The controller records geometric trust-boundary acquisitions, not global-maximum-uncertainty acquisitions.}
\label{fig:sup_111_trust_selection}
\end{figure}

\begin{figure}[!htbp]
\centering
\includegraphics[width=\linewidth]{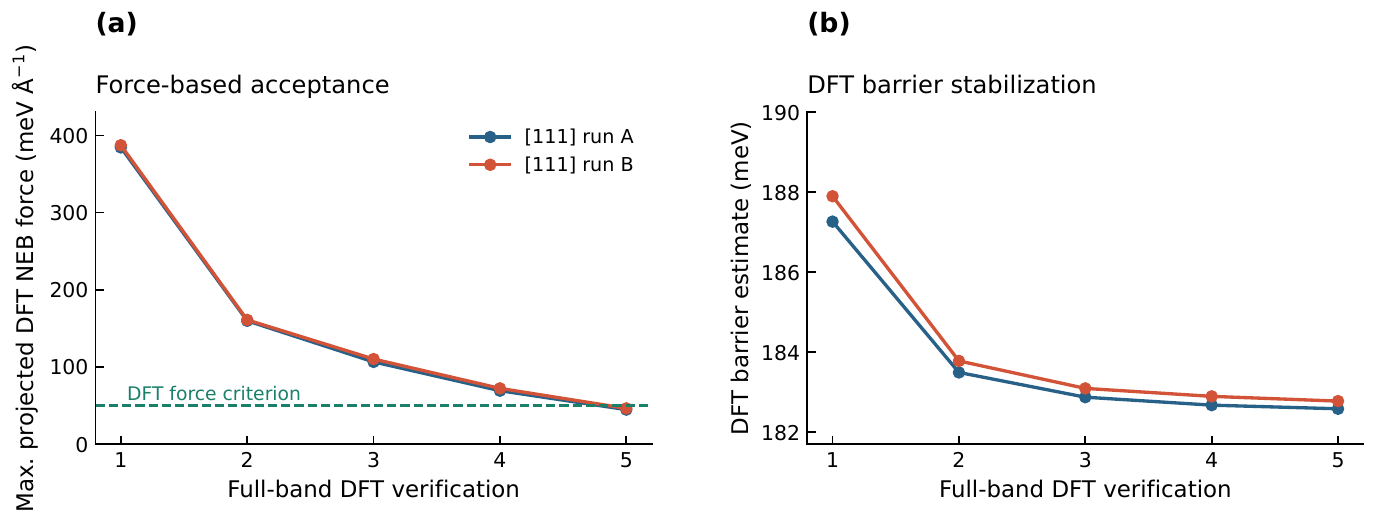}
\caption{Five full-band DFT verification cycles for two closely matched 324-atom, 13-image $[111]$ calculations. (a) Maximum projected DFT NEB force, with the 0.050~eV/\AA{} acceptance threshold shown as a dashed line. (b) Corresponding DFT-evaluated band barriers. Both runs first satisfy the DFT force criterion at the fifth check, yielding barriers of 0.182583 and 0.182775~eV. The barriers from earlier checks are unaccepted intermediate estimates.}
\label{fig:sup_111_verification_repeat}
\end{figure}

\FloatBarrier
\printbibliography[title={References}]